\documentclass[fleqn,usenatbib]{mnras}

\usepackage[T1]{fontenc}

\DeclareRobustCommand{\VAN}[3]{#2}
\let\VANthebibliography\thebibliography
\def\thebibliography{\DeclareRobustCommand{\VAN}[3]{##3}\VANthebibliography}

\usepackage{amsmath,amssymb,bm}
\usepackage{physics}
\usepackage{graphicx}
\usepackage{hyperref}
\usepackage{newtxtext,newtxmath}

\title[Accretion-powered BH-disc collision flares]{Accretion-powered flares from black hole--disc collisions in galactic nuclei}
\author[J. Pelle et al.]{
Joaquin Pelle,$^{1}$\thanks{E-mail: joaquin.pelle@aei.mpg.de}
Kyohei Kawaguchi,$^{1,2}$
Masaru Shibata$^{1,2}$
and Alan Tsz-Lok Lam$^{3,4,1}$
\\
$^{1}$Max-Planck-Institut f\"ur Gravitationsphysik (Albert-Einstein-Institut), Am M\"uhlenberg 1, D-14476 Potsdam-Golm, Germany\\
$^{2}$Center of Gravitational Physics and Quantum Information, Yukawa Institute for Theoretical Physics, Kyoto University, Kyoto, 606-8502, Japan\\
$^{3}$Institute for Gravitation and the Cosmos, The Pennsylvania State University, University Park, PA 16802, USA \\
$^{4}$Department of Physics, The Pennsylvania State University, University Park, PA 16802, USA
}

\date{Accepted XXX. Received YYY; in original form ZZZ}

\pubyear{2026}

\begin{document}
\label{firstpage}
\pagerange{\pageref{firstpage}--\pageref{lastpage}}
\maketitle

\begin{abstract}

Black hole impacts on accretion discs in galactic nuclei can power luminous transients, but predicting their observable signatures is challenging because the post-collision flow is highly time-dependent and inhomogeneous. We present a radiative post-processing framework for relativistic hydrodynamics simulations of black hole--disc collisions. Using physically motivated prescriptions for shock heating, optical depth via an eikonal solver, and photon escape fractions that account for photon trapping and diffusion, we predict light curves and spectral energy distributions over a range of disc densities and collision velocities. Our results indicate that the emission is dominated by the long-lived accretion flow onto the secondary black hole, whose mass supply is highly super-Eddington, rather than by cooling of the unbound ejecta. In the parameter range explored, the luminosity is typically of the order of the Eddington luminosity of the secondary and can reach several times this value, and the emission is generically dominated by soft X-rays. We find that lower velocity collisions produce brighter flares, while the disc surface density mainly controls spectral evolution: low-density discs typically produce $\sim$keV-peaked flares with weak spectral evolution, whereas high-density discs show softer early emission and late-time hardening. A depletion-time estimate calibrated to our results suggests characteristic durations of hours to days for intermediate-mass secondaries, and yields $t_{\rm flare} \propto P_{\rm QPE}$. We discuss implications for QPE-like transients and for the SMBH-binary candidate OJ~287. Our framework does not include radiative feedback on the gas, therefore the luminosities should be interpreted only as baseline estimates of the available power for radiation.

\end{abstract}

\begin{keywords}
accretion, accretion discs -- black hole physics -- hydrodynamics -- radiative transfer -- relativistic processes -- X-rays: general
\end{keywords}

\section{Introduction}
Accretion discs around supermassive black holes (SMBHs) are common at galactic centers, particularly during active phases, as in active galactic nuclei \citep[AGNs,][]{shakura1973black, Novikov:1973, urry1995unified, netzer2015revisiting} or following tidal disruption events \citep[TDEs,][]{rees1988tidal,piran2015disk, mummery2020spectral, gezari2021tidal, dai2021physics}. At the same time, dense nuclear environments are expected to host a population of compact objects. These may form within the accretion disc through disc fragmentation and subsequent stellar evolution \citep{goodman2004supermassive, levin2007starbursts, mckernan2012intermediate, stone2017assisted}, or they may be delivered dynamically from the surrounding nuclear star cluster \citep{bahcall1976star, bahcall1977star, alexander2005stellar, hopman2006effect, o2009gravitational, antonini2012secular, antonini2014distribution, rose2022formation}. Interactions between compact objects and accretion discs are therefore a generic outcome of nuclear dynamics and are expected to influence the evolution of the disc and the observational signatures of the nucleus \citep{mckernan2014intermediate, bartos2017rapid, tagawa2020formation}.

One particularly energetic class of interactions arises when a black hole impacts an accretion disc on a disc-crossing orbit \citep{ivanov1998hydrodynamics}. Due to the typically supersonic velocities involved, such encounters drive strong shocks, and compress and heat the disc gas. The collision triggers a transient accretion flow onto the secondary that can supply mass at a highly super-Eddington rate for a long time after the collision \citep{lam2025black}. These physical conditions are favorable for the production of very luminous electromagnetic transients.

The discovery of quasi-periodic eruptions (QPEs, \citealp{miniutti2019nine, giustini2020x, arcodia2021x, arcodia2024more, nicholl2024quasi, chakraborty2021possible, chakraborty2025discovery, quintin2023tormund, arcodia2025srg, bykov2025further, hernandez2025discovery, baldini2026discovery}; see \citealp{wevers2026observations} for a review) has renewed interest in this class of events. QPEs are luminous repeating X-ray flares observed in galactic nuclei hosting SMBHs of $M_c \sim 10^5$--$10^7M_\odot$ \citep{wevers2022host}. They exhibit recurrence periods of hours to days, large-amplitude variability, and flare durations typically $\sim 10$--$20\%$ of their recurrence period. The physical origin of QPEs remains uncertain. Several proposed mechanisms involve repeated interactions between orbiting objects and accretion discs \citep{dai2010quasi,xian2021x, linial2023emri+, linial2024ultraviolet, franchini2023quasi, tagawa2023flares,linial2025qpes, vurm2025radiation, yao2025star, huang2025multiband, dodd2025perturbing,guo2026testing, suzuguchi2026quasi, jankovivc2026radiation,jankovivc2026radiationB,huang2026resolving, zhu2026unlocking, yao2026star, mondek2026evolution, wang2026quasi, chen2026qpes}. Other models include accretion-disc instabilities or oscillations \citep{sniegowska2020possible, raj2021disk, pan2022disk,kaur2023magnetically, middleton2025quasi}, recurrent mass transfer from an orbiting companion \citep{krolik2022quasiperiodic, metzger2022interacting, zhao2022quasi, king2022quasi}, and self-lensing in SMBH binaries \citep{ingram2021self}. Black hole--disc collisions are a member of the broader class of orbiter-disc interaction models, and may, in principle, contribute to QPE-like phenomenology.

Black hole--disc impacts have also been invoked in the long-standing interpretation of the blazar OJ~287, whose quasi-periodic optical/UV outbursts motivated binary-SMBH scenarios in which a secondary periodically impacts the primary disc \citep{sillanpaa1988oj, lehto1996oj, valtonen2008massive}. Recent analyses have challenged the canonical disc-impact model by the low accretion-disc luminosity inferred from quiescent-state observations \citep{komossa2023absence}, while subsequent work has argued that the interpretation did not account for how accretion power is distributed between thermal disc emission and jet production \citep{valtonen2023need}. Regardless of the ultimate nature of OJ~287, predicting the outcomes of black hole--disc collisions can help place constraints relevant to this discussion \citep[e.g.,][]{linial2023emri+, ressler2025black, chitan2026long}.

Beyond their importance to electromagnetic transients, black hole--disc interactions are also of interest in a multi-messenger context. Compact objects on disc-crossing orbits are expected as extreme-mass-ratio inspiral (EMRI) systems \citep{amaro2007astrophysics, amaro2018relativistic}, which are key targets for future space-based gravitational-wave observatories like LISA and TianQin \citep{amaro2017laser, babak2017science, luo2016tianqin}. In such environments, disc interactions can influence orbital evolution, modulate accretion onto the central object, and potentially generate electromagnetic counterparts to gravitational waves \citep{kocsis2011observable, mckernan2012intermediate, bartos2017rapid, yang2019hierarchical, speri2023probing,kejriwal2024repeating, duque2025constraining, lui2025gravitational,zhan2026hubble,suzuguchi2026possibility}. Even if individual electromagnetic sources are not directly detectable as gravitational-wave sources (e.g.\ because their characteristic frequencies fall outside the sensitive observational range), constraining the rate and properties of disc interactions can inform the underlying compact-object population, thus aiding the interpretation of the observable EMRI subset \citep{chen2022milli, zhou2024probing, zhou2024probingII, chakraborty2025prospects}.

Several studies have investigated aspects of black hole interactions with accretion discs, including analytical estimates \citep{lehto1996oj,pihajoki2016black,linial2023emri+, franchini2023quasi, mummery2025collisions} and (magneto)hydrodynamics simulations of the resulting flows \citep{ivanov1998hydrodynamics, sukova2021stellar, pasham2024case, ressler2024black, ressler2025black, lam2025black, rowan2025hydrodynamic, chitan2026long, liu2026quasi, jiang2026electromagnetic, garain2026massive}. However, predicting their observable electromagnetic signatures remains challenging. Fully coupled radiation-hydrodynamics simulations \citep[e.g.,][]{ohsuga2011global,jiang2014global,skadowski2017radiative} are computationally expensive and limited in parameter coverage, while purely analytic estimates struggle to capture the complex, time-dependent structure of the post-collision flow. Consequently, predictions for observable light curves and spectral energy distributions (SEDs) remain limited.

We investigate the electromagnetic signatures of black hole--disc collisions by radiative post-processing of relativistic hydrodynamics simulations. Using simulation data from \citet[][hereafter, L25]{lam2025black}, we estimate local energy dissipation due to shock heating, photon escape fractions, and emergent spectra, while accounting for radiative diffusion and photon trapping. This approach allows us to compute light curves and SEDs over a range of orbital and disc parameters.

Previous models of black hole--disc collisions have generally focused on the emission by cooling of the unbound ejecta produced during the disc impact. Our results, however, suggest that, for a broad region of parameter space, the dominant source of radiation is instead the highly super-Eddington mass accretion flow \citep[e.g.,][]{inayoshi2016hyper, kitaki2018systematic, kitaki2021origins, toyouchi2024radiation} that develops around the secondary black hole after the collision, producing long-lived soft X-ray flares.

This paper is organized as follows. In Sec.~\ref{sec:physical-setup} we describe the physical setup and characteristic scales of the black hole--disc collision scenario. Sec.~\ref{sec:framework} summarizes the radiative post-processing framework, including shock heating, optical depth calculations, and emergent SEDs. Sec.~\ref{sec:results} presents the resulting bolometric- and specific-luminosity light curves along with time-averaged SEDs. In Sec.~\ref{sec:discussion}, we discuss astrophysical implications of our results. We conclude in Sec.~\ref{sec:conclusions} with a summary and a discussion of limitations and prospects for future work. Throughout this paper, $G$, $c$, and $k_{\rm B}$ denote the gravitational constant, speed of light, and Boltzmann's constant, respectively. 

\section{Physical setup}
\label{sec:physical-setup}
We consider the interaction between an orbiting black hole of mass $m_{\rm BH}$ (hereafter, the secondary) and the accretion disc surrounding a supermassive black hole (SMBH) of mass $M_{\rm c}$ (the primary). The secondary crosses the disc with a relative velocity $v$, which depends on the orbital parameters, such as inclination and eccentricity, as well as on the local velocity of the disc material. For reference, when mapping to astrophysical systems in Sec.~\ref{sec:discussion}, we will approximate the relative velocity as the local circular Keplerian velocity, adequate for low-eccentricity EMRIs crossing the disc at generic inclination angles. At this point, we treat $v$ as an independent parameter. 

The disc is characterized locally by its surface density $\Sigma$, vertical scale height $H$, and adiabatic index $\Gamma$. Throughout this work, these quantities are understood as local disc properties evaluated at the orbital radius of the secondary. A characteristic timescale is the disc-crossing time $t_{\rm cross}:= 2H/v$. For convenience, we also define the dimensionless scale height, $h:=H/r_g$, where $r_g:=Gm_{\rm BH}/c^2 \approx 1.47 \times 10^5 (m_{\rm BH}/M_\odot)\,{\rm cm}$ is the gravitational radius of the secondary (see Table~\ref{tab:symbols} for the definition of other relevant quantities).

A useful interaction length scale is the accretion radius of the secondary,
\begin{equation}
\begin{aligned}
r_a &:= \frac{2Gm_{\rm BH}}{v^2} \approx 1.2 \times 10^{12} m_\mathrm{BH,4} v_{0.05}^{-2}~\mathrm{cm},
\end{aligned}
\end{equation}
which sets the characteristic size of the region from which disc material is gravitationally focused toward the secondary during the encounter. Here, $m_\mathrm{BH,4} := m_{\rm BH}/(10^4M_\odot)$, and $v_{0.05} := v/(0.05c)$. Associated with this scale is the characteristic intercepted mass,
\begin{equation}
\begin{aligned}
m_a &:= \pi r_a^2 \Sigma \approx 4.4 \times 10^{28} m_\mathrm{BH,4}^2 v_{0.05}^{-4} \Sigma_4 \,\,{\rm g}, 
\end{aligned}
\end{equation}
and the corresponding mass flux,
\begin{equation}
\begin{aligned}
\dot{m}_a &:= \pi r_a v \Sigma \approx 5.6 \times 10^{25} m_\mathrm{BH,4} v_{0.05}^{-1} \Sigma_4 \,\,{\rm g/s},
\end{aligned}
\end{equation}
where $\Sigma_4 := \Sigma/(10^4~{\rm g~cm^{-2}})$. For comparison, we define the Eddington mass accretion rate of the secondary as $\dot{m}_{\rm Edd,BH} := 10\,L_{\rm Edd,BH} / c^2$, where $L_{\rm Edd,BH} = 4\pi G c m_{\rm BH}/\kappa \approx 1.4 \times 10^{42} m_{\rm BH,4}~{\rm erg~s^{-1}}$ is the Eddington luminosity of the secondary and we assumed a radiative efficiency of $10\%$ and opacity $\kappa \approx 0.34~{\rm cm^2~g^{-1}}$. In terms of this rate, we can write 
\begin{equation}
\dot m_a \approx 3.7 \times 10^3 v_{0.05}^{-1} \Sigma_4 \,\dot m_{\rm Edd,BH}. 
\end{equation}
Therefore, matter can typically be supplied at highly super-Eddington rates. These scalings also suggest that the collision outcome is highly sensitive to the impact velocity. Since $r_a\propto v^{-2}$, the intercepted mass depends steeply on velocity, $m_a\propto v^{-4}$. Thus, slower encounters intercept substantially more gas and supply a larger energy reservoir to feed the post-collision accretion flow, whereas faster encounters are naturally less energetic even if they remain super-Eddington (because $\dot m_a \propto v^{-1}$). As we show in Sec.~\ref{sec:results}, our results confirm these expectations.

For the fluid variables, we use data from the axisymmetric relativistic hydrodynamics simulations in L25, which were performed with the {\tt SACRA-2D} code \citep{lam2025new}. In those simulations, the Euler equations are solved in the frame of the secondary, and the disc is initially assumed to move with uniform vertical velocity towards the secondary in that frame. Prior to the encounter, the disc is modeled locally as a horizontally homogeneous fluid slab in vertical hydrostatic equilibrium under the gravity of the primary (which is not in the simulation domain). The slab extends across the numerical domain, whose cylindrical radius is $12800\,r_g$. The relevant condition for this local approximation is that $r_a$ be small compared with the horizontal scale of variation of the disc properties. Taking the orbital radius $R$ as a characteristic variation scale and adopting $v\simeq v_{\rm K}=(GM_{\rm c}/R)^{1/2}$, we obtain $r_a\simeq 2Gm_{\rm BH}/v_{\rm K}^2=2qR$, where $q:=m_{\rm BH}/M_{\rm c}$ is the mass ratio of the system. Thus, for $q\ll1$, the local approximation and horizontal homogeneity are reasonable assumptions. The fluid is assumed to follow a $\Gamma$-law equation of state with $\Gamma = 4/3$, corresponding to the radiation-pressure dominated regime. The vertical gravitational field due to the primary is approximated as linear in height above the midplane, with a phenomenological cutoff at a height $z_{\rm cut}$. For more details on the simulation setup, we refer the reader to L25.

Several assumptions underlie our post-processing approach. First, we neglect radiative feedback on the gas dynamics. This is not because radiation forces are necessarily dynamically unimportant, but it constitutes a limitation of our present work. Especially in the most luminous models, radiation forces and cooling could modify the structure of the post-collision flow, reduce the net accretion rate, and change the emergent luminosity. The aim of our calculation is mainly to estimate how much power is made available by the collision, where it is produced as the flow evolves, and how it would be processed into observable radiation.  A self-consistent assessment of radiative effects requires radiation-hydrodynamics simulations, which are beyond the scope of this work. In this sense, the luminosities reported below should be interpreted as baseline estimates of the radiative power available in the hydrodynamic flow, rather than as self-consistent predictions of a radiation-regulated accretion flow. Second, even though the simulations are performed in general relativity (GR), we do not include GR corrections in the radiative calculations. This simplification should be acceptable because the regions that dominate the observable emission are typically located at distances from the secondary where GR effects are subdominant. The innermost regions close to the event horizon, where strong-field effects are important, are largely hidden from view by the high optical depth.

\begin{table}
\centering
\caption{Main symbols used in this work.}
\begin{tabular}{ll}
\hline
Symbol & Meaning \\
\hline
$M_{\mathrm c}$ & Mass of the central SMBH (primary) \\
$m_{\mathrm{BH}}$ & Mass of the impactor BH (secondary) \\
$r_g$ & Gravitational radius of the secondary, $Gm_{\mathrm{BH}}/c^2$ \\
$t_g$ & Gravitational time of the secondary, $Gm_{\mathrm{BH}}/c^3$ \\
$v$ & Relative velocity of the secondary through the disc \\
$\Sigma$ & disc surface density at the crossing radius \\
$H$ & disc scale height at the crossing radius \\
$h$ & Dimensionless disc scale height, $H/r_g$ \\
$r_a$ & Accretion radius, $2Gm_{\rm BH}/v^2$  \\ 
$m_a$ & Characteristic intercepted mass, $\pi r_a^2 \Sigma$ \\ 
$\dot m_a$ & Characteristic mass flux, $\pi r_a v \Sigma$ \\
$t_{\mathrm{cross}}$ & disc crossing time, $2H/v$ \\
$L_{\mathrm{Edd,BH}}$ & Eddington luminosity of the secondary black hole \\
$R$ & Orbital radius of the secondary around the primary \\
$P_{\mathrm{orb}}$ & Orbital period of the secondary around the primary \\
$P_{\mathrm{QPE}}$ & Fiducial QPE recurrence period, $P_{\mathrm{orb}}/2$ \\
\hline
\end{tabular}
\label{tab:symbols}
\end{table}

\section{Radiative post-processing framework}
\label{sec:framework}
We map hydrodynamic snapshot data to estimates of electromagnetic observables by volume integrals of the form:
\begin{equation}
L_{\mathrm{bol}}(t) \;\approx\; \int \dot{e}(\bm{x},t)\, f_{\mathrm{esc}}(\bm{x},t)\, \dd V ,
\label{eq:Lbol_master}
\end{equation}
where $L_{\mathrm{bol}}$ is the bolometric luminosity, $\dot{e}$ is the local photon energy generation rate density and $f_{\mathrm{esc}}$ is an effective escape fraction capturing photon trapping and finite diffusion time effects, and the integration is done over the simulation domain, covering $x \leq L$ and $|z| \leq L$, where $(x,z)$ are the cylindrical radius and vertical coordinate centered on the secondary, respectively, and $L=12800\,r_g$. For the specific luminosities, we add a local spectral-weighting factor to the integrand in a manner we detail further below.

\subsection{Local internal energy generation rate density \texorpdfstring{$\dot{e}$}{e-dot}}
\label{sec:edot}
We compute $\dot{e}$ based on a shock-heating estimator. For a planar shock of area $A$ and shock speed $v_{\mathrm{sh}}$, the shock-heating rate is obtained as the mass flux across the shock multiplied by the jump in specific internal energy, as follows \citep[e.g.,][]{landau1987fluid}: 
\begin{equation}
\dot{E} = A\,\rho_u \,(v_u - v_{\mathrm{sh}})\,(\varepsilon_d - \varepsilon_u),
\label{eq:shock_Edot}
\end{equation}
where $\rho$, $\varepsilon$, and $v$ are the density, specific internal energy, and velocity in the frame of the secondary, respectively, and subscripts $u$ and $d$ denote upstream and downstream states. In discrete form, across a cell interface of area $\Delta A$, assigned to a downstream cell volume $\Delta V$, we define
\begin{equation}
\dot{e} \,\Delta V = \Delta A\,\rho_u \,(v_u - v_{\mathrm{sh}})\,(\varepsilon_d - \varepsilon_u).
\label{eq:shock_edot}
\end{equation}
We refer the reader to Appendix~\ref{app:shock_id} for details on the shock identification criteria and shock speed calculation. 

\subsection{Optical depth}
\label{sec:tau}
The photon escape fraction is tied to the optical depth in the medium. In a scattering-dominated, optically thick medium, photons preferentially escape along paths of least integrated optical depth. To model photon escape in such a medium, we compute the optical depth field $\tau(\bm{x})$ as a minimum over paths between $\bm{x}$ and the boundary of the domain $\partial\Omega$,
\begin{equation}
\tau(\bm{x}) = \min_{\gamma \in \mathcal{C}(\bm{x})} \int_\gamma \kappa\rho\,\dd l,
\label{eq:tau_def}
\end{equation}
where $\mathcal{C}(\bm{x})$ is the set of smooth paths starting from any point in $\partial \Omega$ and ending in $\bm{x}$. This choice is motivated by the fact that, in strongly inhomogeneous and anisotropic flows such as those following a black hole--disc collision, the minimum-escape paths are generically not along a fixed geometric direction (e.g., radial or vertical as often assumed). Therefore, an optical depth defined along, for example, radial rays would not be appropriate to capture the true escape fraction, as it may artificially block regions of the gas in a highly non-spherical geometry. This approach is similar to that used in some neutrino-leakage schemes \citep[e.g.,][]{perego2014neutrino}.

The global optimization problem as formulated in Eq.~\eqref{eq:tau_def} is not suitable for direct solution. Instead, we rely on the property that the desired solution satisfies an eikonal equation,
\begin{equation}
\abs{\nabla\tau} = \kappa\rho,\qquad \tau|_{\partial\Omega}=0,
\label{eq:eikonal}
\end{equation}
which we solve using a fast-sweeping method \citep{zhao2005fast} implemented in the Julia package Eikonal.jl, with modifications to enable the multi-block grids on which our hydrodynamic data is given. From each cell, we recover minimum-escape paths $\gamma_{\mathrm{esc}}$ and path lengths $\ell_{\mathrm{esc}}(\bm{x})$ by gradient descent along $-\nabla\tau$.

We adopt a constant opacity $\kappa=0.34~\mathrm{cm^2\,g^{-1}}$ for the optical-depth computation, consistent with electron scattering in a fully ionized gas with cosmic composition, the dominant opacity in the temperature and density range relevant for the post-collision flow. True absorption processes (mainly free--free) are subdominant for the total optical depth but remain important for photon production and thermalization. These effects intervene in the photon temperature estimate described in Sec.~\ref{sec:results_temp}. We have verified that true absorption does not significantly attenuate the escaping radiation, and therefore use only the electron-scattering opacity for the escape fraction (see Sec.~\ref{sec:results_temp} for more details). 

\subsection{Escape fraction}
\label{sec:fesc}
In optically thick flows, radiation can be trapped when inward advection is faster than outward diffusion \cite{begelman1978black}. To account approximately for photon trapping and diffusion effects, we adopt the phenomenological prescription
\begin{equation}
f_{\mathrm{esc}}(\bm{x},t) := f_{\mathrm{adv}}(\bm{x},t)\, f_{\mathrm{diff}}(\bm{x},t),
\label{eq:fesc_factorized}
\end{equation}
representing suppression by photon trapping and by finite diffusion time, respectively.

\subsubsection{Photon trapping}
\label{sec:advection_trapping}
Our photon-trapping factor is motivated by an analysis of the advection timescale of photons on an underlying hydrodynamic profile (see Appendix~\ref{app:advection_trapping} for the details). The factor is locally defined along the corresponding minimum-escape path $\gamma_{\mathrm{esc}}$ as:
\begin{equation}
f_{\mathrm{adv}} :=
\exp\!\left[\frac{4}{c}\int_{\gamma_{\mathrm{esc}}} v^-\,\dd\tau\right],
\qquad
v^- := \min(v_r,0),
\label{eq:fadv_main}
\end{equation}
where $v_r$ is the radial velocity component and $\dd\tau=\kappa\rho\,\dd \ell$ is the differential optical depth along
$\gamma_{\mathrm{esc}}$. This prescription attenuates radiation generated in regions where large optical depth and inward advection would tend to drive photons inward, thereby suppressing photon escape.

\subsubsection{Finite diffusion-time suppression}
We estimate the diffusion time as
\begin{equation}
t_{\mathrm{diff}}(\bm{x},t) := \tau(\bm{x},t)\,\ell_{\mathrm{esc}}(\bm{x},t) / c ,
\label{eq:tdiff}
\end{equation}
and apply an attenuation relative to the time elapsed from a reference collision time $t_{\mathrm{coll}}=t_{\rm cross}:=2H/v$,
\begin{equation}
f_{\mathrm{diff}} :=
\exp\!\left[-\frac{t_{\mathrm{diff}}}{t-t_{\mathrm{coll}}}\right].
\label{eq:fdiff}
\end{equation}
Equations \eqref{eq:fadv_main}--\eqref{eq:fdiff} define the escape fraction used in Eq.~\eqref{eq:Lbol_master}.

\subsection{Local photon temperature}

To estimate specific luminosities at given frequencies, we need to assign a characteristic spectrum to the radiation escaping from each region of the flow. We characterize the local radiation field by a characteristic photon temperature $T$, defined through 
\begin{equation}
\frac{e_\gamma}{n_\gamma} \equiv 3 k_{\rm B}T ,
\end{equation}
where $e_\gamma$ is the radiation energy density, and $n_\gamma$ is the photon number density. The hydrodynamics simulations provide the internal energy density, $\rho \varepsilon$, which is assumed to be dominated by radiation, so that $e_\gamma \approx \rho \varepsilon$. However, they do not provide $n_\gamma$. We therefore estimate $n_\gamma$, and hence $T$, following a thermalization argument from \citet{nakar2010early}. 

We consider photon production by free--free emission, which is the relevant thermal continuum process for a fully ionized gas with cosmic composition at temperatures $\gtrsim 10^5~$K. We further assume local Compton equilibrium of electrons with the radiation field, which is expected to hold in optically thick regions, where electron scattering provides sufficient coupling between the species\footnote{This is not valid in optically thin regions. However, we have verified that these regions contribute negligibly to the bolometric luminosity, and therefore focus only on radiation produced where strong Compton coupling holds.}. Under this assumption, the electron temperature follows the photon temperature, $T_e \approx T$. This does not imply full thermodynamic equilibrium: scattering redistributes energy between electrons and photons, but it conserves photon number. The electrons may therefore be Compton-cooled while the radiation field remains photon-starved, if true emission does not produce enough photons before escape. An important question is then whether photon production is sufficient to supply the blackbody (thermal-equilibrium) number density, $n_{\rm BB} \approx a_{\rm rad} T_{\rm BB}^4 / 3k_{\rm B} T_{\rm BB}$, where $T_{\rm BB}:=\left(e_\gamma/a_{\rm rad}\right)^{1/4}$ is the blackbody temperature for the given energy density. If not, the same total energy is shared among a smaller number of photons, and therefore $T>T_{\rm BB}$, yielding a harder spectrum than would be obtained in thermal equilibrium. This case corresponds to the so-called photon-starved regime. 

We approximate the characteristic time photons reside in the flow as $t_{\rm res} \approx t_{\rm diff}$\footnote{Strictly speaking, one could replace $t_\mathrm{diff}$ by $\min(t-t_{\rm coll},\, t_\mathrm{diff})$. We have verified that this modification does not affect the results, so we keep the simpler expression.}, in which case the photon number density produced during that time can be approximated as
\begin{equation}
n_\gamma \approx t_{\rm diff}\,\dot n_{\rm ff}(\rho,T)\,\xi(T),
\end{equation}
where $\dot n_{\rm ff} \approx 3 \times 10^{36}~{\rm cm^{-3}\,s^{-1}}(\rho/{\rm g \,cm^{-3}})^2(T/K)^{-1/2}$ is the free--free production rate of photons per logarithmic energy interval at $3k_{\rm B}T$, and $\xi(T)$ is a correction factor accounting for lower-energy free--free photons that can be Compton upscattered to $\sim 3k_{\rm B}T$ \citep[for the expression of $\xi(T)$, see][]{nakar2010early}. Using $n_\gamma \approx \rho\varepsilon/ 3k_{\rm B}T$, we obtain an implicit equation for $T$,
\begin{equation}
\frac{\rho\varepsilon}{3k_{\rm B} T}
=
t_{\rm diff}\dot n_{\rm ff}(\rho,T)\xi(T).
\label{eq:T_implicit}
\end{equation}
We denote the solution to this equation as $T_{\rm ff}$. If $T_{\rm ff} > T_{\rm BB}$, that means photon production is insufficient to reach the blackbody photon number before escape, corresponding to the photon-starved regime. In that case, we set $T=T_{\rm ff}$, up to the upper bounds discussed below. If instead $T_{\rm ff} \leq T_{\rm BB}$, free--free emission and absorption are sufficient to drive the radiation toward balance, and the photon number saturates at the blackbody value. In that case, we set $T=T_{\rm BB}$. 

We further constrain the photon temperature thus obtained by accounting for pair-production self-regulation. At sufficiently high photon temperatures, photon--photon interactions efficiently produce electron--positron pairs, increasing the lepton density and degrading the photon energy. This regulates the photon temperature to remain below $T_{\gamma\gamma} \sim 100\text{--}200~\mathrm{keV}$ \citep{weaver1976structure, katz2010fast}. Therefore, we cap the photon temperature at $T_{\gamma\gamma} = 200~\mathrm{keV}$ to account for this self-regulation in a phenomenological manner. In summary, the photon temperature assigned to each cell is defined as $T = \min\!\left(\max\!\left(T_{\mathrm{BB}},\,T_{\rm ff}\right),\,T_{\gamma\gamma}\right)$.

We then check a posteriori whether this assignment is consistent with the assumption that the internal energy density is carried predominantly by radiation. If we assume Coulomb coupling is strong enough to thermalize electrons and ions ($T_i \approx T_e \approx T$), which is valid in the dense, optically thick regions, their combined internal energy density would be
$e_{\rm gas} \approx {3 \over 2} (n_i + n_e) k_{\rm B} T \approx {3\over 2} {\rho \over {\mu m_p}} k_{\rm B} T$, with $\mu$ the mean molecular weight per particle. The radiation-dominated assumption requires $e_{\rm gas} \ll \rho \varepsilon$. If this condition is not satisfied, the energy budget underlying Eq.~\eqref{eq:T_implicit} is not valid. In that case, more detailed considerations of the advection and cooling timescales would be required to determine the fraction of internal energy carried by photons. However, we have checked that regions where this condition is not satisfied contribute negligibly to the escaping radiation. We therefore disregard them in the spectral calculations.

Finally, we have verified that considering free--free absorption in the optical depth does not affect the results. Defining a free--free absorption opacity as $\kappa_{\rm abs} \sim \dot n_{\rm ff}(\rho,T_{\rm BB})/(c\rho \,n_{\rm BB}(T_{\rm BB}))$, one may estimate an absorption optical depth $\tau_{\rm abs}\sim \kappa_{\rm abs}\rho \ell_{\rm esc}$. The corresponding effective absorption optical depth is then $\tau_{\rm eff}\sim \sqrt{\tau_{\rm abs}\tau}$, where $\tau$ is the scattering optical depth defined in Section~\ref{sec:tau}. In the scattering-dominated regime relevant here, we find $\tau_{\rm eff}\ll\tau$, and including an attenuation factor $\exp(-\tau_{\rm eff})$ does not produce significant changes in the light curves. This supports our use of scattering optical depth alone in the escape fraction.

\subsection{Specific luminosities}
\label{sec:specific_luminosities}
Once the local temperature map throughout the flow has been obtained, we define rough estimates of the specific luminosity at a given frequency $\nu$ in volume-integral form as
\begin{equation}
\nu L_\nu \approx
\nu \int \dot{e}f_\mathrm{esc}\,
\frac{\pi B_\nu(T)}{\sigma T^4}\,\dd V,
\label{eq:nuLnu_proxy}
\end{equation}
where $\sigma=a_\mathrm{rad}c/4$ is the Stefan-Boltzmann constant, $B_\nu$ is the Planck function, and the integral is restricted to the optically thick, photon-dominated region where the assumptions underlying our temperature calculation are satisfied. Note that the weighting factor implicitly assumes that photons are thermally distributed at temperature $T$, which is not necessarily true. This spectral weighting should therefore be interpreted as a proxy based on characteristic photon energies, rather than an exact spectral shape. Alternatively, we could use a spectral weight proportional to the free--free emissivity. We have verified that this produces a modest broadening of the SED, particularly enhancing emission at lower photon energies (e.g., $\sim 100$\,eV) relative to the $\sim 1$\,keV peak. This does not affect the qualitative trends and our main conclusions, and we therefore retain the blackbody weighting in Eq.~\eqref{eq:nuLnu_proxy} for simplicity.

\subsection{Limitations of the present framework}
Several limitations should be kept in mind when interpreting our results. First, the hydrodynamics simulations neglect radiation forces and radiative cooling, which could modify the gas dynamics, in both optically thick and optically thin regions. Radiation-driven outflows, as found in super-Eddington accretion simulations \citep{ohsuga2011global}, can reduce the accretion rate and the associated energy release. On the other hand, a lower inflow density and outward transport of radiation can reduce inward photon trapping and increase the fraction of radiation that escapes. Photon trapping can nevertheless remain important in the dense inflow, while radiation escapes more readily near the disc surface \citep{2016MNRAS.456.3929S}. In the BHL geometry, \citet{zanotti2011general} found reduced accretion rates compared with purely hydrodynamic evolutions, while retaining luminosities of a few times the Eddington luminosity. These competing effects prevent us from assigning a definite direction to the luminosity correction in our collision problem. Radiation-hydrodynamics simulations will be required to quantify these effects more precisely in the future. Second, our radiative post-processing framework approximates photon escape using a minimum-escape optical-depth prescription and therefore does not capture the angular dependence of the emergent radiation field. Third, the simulations are local, i.e., we focus on the local interaction between the secondary black hole and the accretion disc around the primary, and do not include the global structure of the disc or the orbital evolution of the secondary.

Despite these simplifications, the main conclusions of this work rely primarily on robust hydrodynamic features of the collision, namely the formation of large-scale shocks and the development of a long-lived super-Eddington mass accretion flow. We expect these features to persist at least qualitatively in more realistic radiation-hydrodynamics simulations, although radiative feedback may change the accretion rate, the degree of photon trapping, and the emergent luminosity.

\section{Results}
\label{sec:results}
\subsection{Parameter space and scalings}
\label{sec:parameter_space}
The simulations analyzed here are drawn from the suite presented in L25. Since fluid self-gravity and radiative feedback are neglected, the hydrodynamic evolution is invariant under a constant rescaling of the gas density for the assumed $\Gamma$-law equation of state. As a result, the hydrodynamics simulations in L25 are characterized primarily by the collision velocity $v$ and the dimensionless disc thickness $H/r_g$\footnote{This ratio should not be confused with the vertical aspect ratio of the disc, $H/R$, where $R$ is the crossing (orbital) radius. Since the simulations are local, they are agnostic with respect to $R$, which we only specify when mapping the results to astrophysical systems in Sec.~\ref{sec:discussion}.}.

Although the original simulation suite spans a broader range of velocities, in this work we focus on a representative subset of models with collision velocities $v = 0.05c,\,0.1c$, and $0.2c$, which samples different velocity regimes. The simulations also include several values of the disc scale height, $H = 200$, $500$, and $1000\,r_g$. Within this range the qualitative properties of the post-collision flow depend only weakly on $H/r_g$ compared to their dependence on $v$, so for clarity the analysis presented below focuses on the $H=1000\,r_g$ models. We show below that the relevant emission timescales are much longer than the disc-crossing time, so $H$ is expected to impact at most the earliest stages of the evolution.

In contrast to the underlying hydrodynamics solution, the radiative post-processing is not invariant under a density rescaling, because the optical depth, diffusion time, photon escape fraction, and spectral properties depend explicitly on the physical density normalization or, equivalently, on $\Sigma$. We therefore assign physical values to $\Sigma$ when computing radiative diagnostics on top of the simulation data. In this work we consider $\Sigma = 10^3,\,10^4,\,10^5$, and $10^6~{\rm g\,cm^{-2}}$, which span a broad range expected for accretion discs in galactic nuclei at radii relevant for nuclear transients.

The secondary mass $m_{\rm BH}$ mainly sets the physical normalization of the problem. At fixed collision velocity $v$ and dimensionless disc thickness $H/r_g$, one has $\rho \propto \Sigma/m_{\rm BH}$  (since $r_g\propto m_{\rm BH}$), $\dot e \propto \Sigma/m_{\rm BH}^2$, $\tau \sim \kappa \rho\,\ell_{\rm esc} \propto \Sigma$ (i.e. the optical depth at a given position in units of $r_g$ is independent of $m_{\rm BH}$), and $t_{\rm diff} \sim \tau \ell_{\rm esc}/c \propto \Sigma m_{\rm BH}$. Consequently, the bolometric luminosity in Eddington units, $L_{\rm bol}/L_{\rm Edd,BH}$, is independent of $m_{\rm BH}$ for a fixed set of $(v,\Sigma, H/r_g)$. The role of $m_{\rm BH}$ is therefore mainly to set the physical luminosity and timescale normalizations, with only a minor additional effect on the spectrum through the local temperature estimate\footnote{In radiation-hydrodynamics simulations, $m_{\rm BH}$ could enter more directly through the radiative feedback and cooling.}. More precisely, for a fixed set of $(v,\Sigma, H/r_g)$, the implicit temperature solution of Eq.~\eqref{eq:T_implicit} is independent of $m_{\rm BH}$ since $\rho \propto m_{\rm BH}^{-1}$, $t_{\rm diff} \propto m_{\rm BH}$, and $\dot{n}_{\rm ff} \propto \rho^2 \propto m_{\rm BH}^{-2}$. The only dependence of the final temperature estimate on $m_{\rm BH}$ is through the lower bound, $T_{\rm BB} \propto \rho^{1/4} \propto m_{\rm BH}^{-1/4}$. As we show in Sec.~\ref{subsec:sed}, this results in a weak spectral dependence on $m_{\rm BH}$ within the intermediate-mass black hole (IMBH) range.

\subsection{Bolometric emission}

\label{subsec:bolometric_luminosities}
\begin{figure*}
    \centering
    \includegraphics[width=\textwidth]{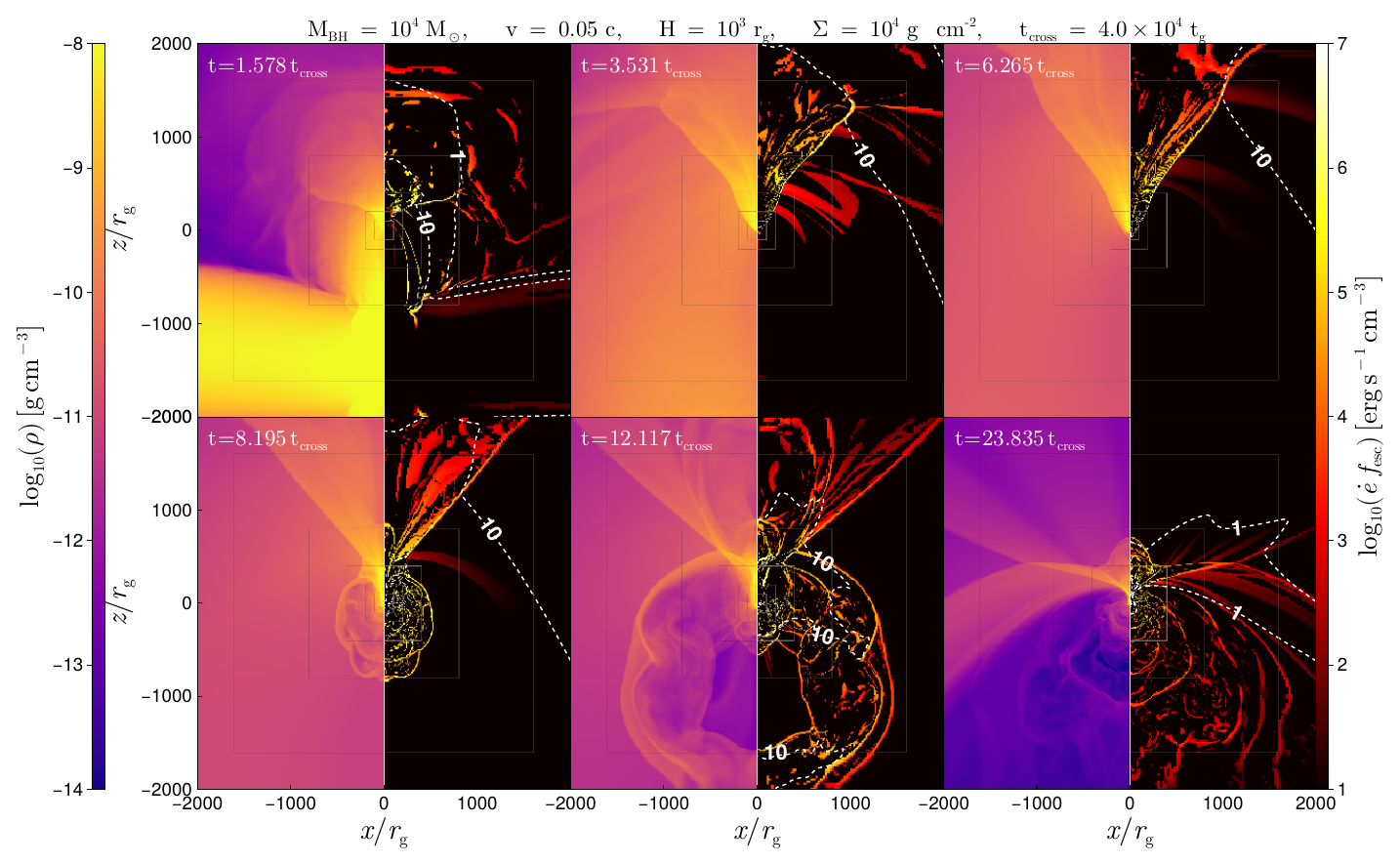}
    \vspace{-3mm}
    \caption{
    \textbf{Rest-mass density and energy-generation-rate density weighted by escape fraction together with optical-depth contours at selected times in $x$-$z$ planes.}
    Snapshots from the $v=0.05c$, $\Sigma=10^4\,\mathrm{g\,cm^{-2}}$ model showing the rest-mass density $\rho$ (left side of each panel) and $\dot e\,f_{\rm esc}$ (right side of each panel). The panels are centered on the secondary, the frame in which the simulations were performed. The disc moves downwards in this frame and is seen only in the top left panel. The normalization of both quantities corresponds to $m_{\rm BH} = 10^4 M_\odot$. Overlaid curves show contours of optical depth $\tau$ computed with the eikonal solver. Each panel corresponds to a different time (as labeled in the figure), illustrating the evolution of the spatial distribution of radiative power that can escape to the observer. 
    }
    \label{fig:escape_heating_tau_contours}
\end{figure*}

\begin{figure*}
    \centering
    \includegraphics[width=\textwidth]{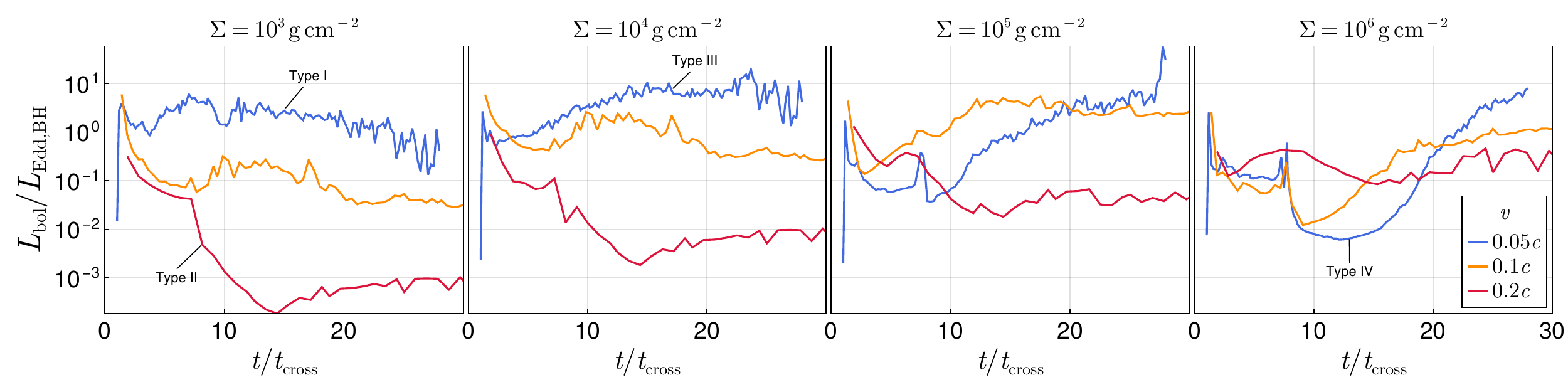}
    \vspace{-3mm}
    \caption{
    \textbf{Bolometric luminosities.} Bolometric luminosity in units of the Eddington luminosity, $L/L_{\rm Edd,BH}$, as a function of the dimensionless time $t/ t_{\rm cross}$, for four values of the disc surface density $\Sigma$. Colors denote the collision velocity $v$, as indicated. Representative examples of the four lightcurve morphology types (Types I–IV) are indicated in the panels. Because the horizontal axis is normalized by $t_{\rm cross}=2H/v$, the same extent in $t/t_{\rm cross}$ corresponds to a longer physical duration for lower collision velocities. Short-timescale fluctuations are unlikely to correspond to observable variability and should not be overinterpreted.}
    \label{fig:bolometric_luminosity_H1000}
\end{figure*}

\subsubsection{Flow morphology and emitting regions}

Before discussing the integrated light curves, it is useful to identify the main emitting regions in the flow. Qualitatively, the post-impact evolution can be separated broadly into two stages. Shortly after the black hole crosses and emerges from the disc, the emission is dominated by shocked gas associated with a Bondi--Hoyle--Lyttleton (BHL) shock \citep{hoyle1939effect,bondi1944mechanism,edgar2004review}. This is the first stage. This BHL flow structure forms as disc material is accelerated by large pressure gradients toward the secondary faster than the secondary recedes from the disc (see the first three time slices in Fig.~\ref{fig:escape_heating_tau_contours}). At later times (see the later three time slices in Fig.~\ref{fig:escape_heating_tau_contours}), this BHL shock contribution weakens, while a broader region in the wake of the secondary becomes increasingly important. This later region is fed by gas that is driven toward the secondary and shocked, but is not accreted. Instead, it is deflected around the black hole, reaccelerated downwards, and injected back into the wake, where it collides with newly incoming gas and previously shocked layers. This process builds a hot, turbulent bubble behind the secondary, and convective motion is driven within it. Photons are then emitted mainly by shocked structures embedded in this region. This two-stage description is only approximate, since both components can coexist for part of the evolution, but it provides a useful guide to the flow morphology shown in Fig.~\ref{fig:escape_heating_tau_contours}.

Fig.~\ref{fig:escape_heating_tau_contours} illustrates this spatial evolution for a representative model with $\Sigma = 10^4~{\rm g~cm^{-2}}$ and $v=0.05c$, with units corresponding to $m_{\rm BH}=10^4M_\odot$. Each pair of panels shows one simulation snapshot in the frame comoving with the secondary black hole. The left half of each snapshot shows the rest-mass density, while the right half shows the local escaping bolometric emissivity, $\dot e f_{\rm esc}$. The emissivity maps are intrinsically more filamentary than the density maps because the heating rate $\dot e$ is evaluated through our shock-identification prescription. As a result, significant emission is produced in cells identified as shocked, or in their immediate surroundings, rather than throughout the diffuse flow. The white dashed curves overlaid on the emissivity maps in the right half of each panel show optical-depth contours, $\tau=1$ and $10$, with the corresponding values indicated by labels on the contours. In some snapshots, some contours do not appear because their levels lie outside the displayed region. 

For interpreting the figure, it is important to keep in mind that they are set in the frame of the secondary black hole. In this frame, the secondary is fixed at the origin and the unperturbed disc material moves downward through the computational domain. Consequently, shortly after the collision the dense disc layer occupies the lower part of the displayed region, but later moves out of the plotted window. A wider view of the flow evolution can be seen in L25. Here we show only a more immediate neighborhood of the secondary, where most of the emission is generated.

The first snapshot, at $t \approx 1.5\,t_{\rm cross}$, shows the black hole shortly after emerging from the dense disc. The escaping emissivity is concentrated in narrow shocked structures close to the black hole and along the BHL shock. At $t \approx 3.5\,t_{\rm cross}$ and $t \approx 6.2\,t_{\rm cross}$, the dense disc material has largely moved downward out of the plotted region, while captured material remains around the secondary. These snapshots correspond to the early, BHL-shock-dominated stage described above.

At later times, $t \approx 8.2\,t_{\rm cross}$, $12\,t_{\rm cross}$, and $23\,t_{\rm cross}$, the initially more coherent shock structure has weakened, and the emission is instead associated with the broader turbulent, convective region around the secondary. The emissivity becomes more spatially extended, with bright filaments tracing shocks generated within this turbulent bubble. These later panels illustrate the second stage of the evolution, in which the escaping radiation is mainly powered by the convective region around the secondary. 

Considering $2\pi R \,\dot ef_{\rm esc}$ instead, i.e. weighting the emission by the cylindrical volume element, does not qualitatively change the picture. The colormap still traces the same shocked structures, with only a modest reweighting toward larger cylindrical radii.

\subsubsection{Bolometric light curves}

Fig.~\ref{fig:bolometric_luminosity_H1000} shows the bolometric light curves. The accretion-powered emission typically persists for a substantial fraction of the simulated interval, of order $\sim 10$--$30\,t_{\rm cross}$ depending on the model. For reference, the disc-crossing time scales as $t_{\rm cross}= 4\times10^4 \,t_g \, v_{0.05}^{-1}h_3$, where $h_3:=H/10^3r_g$. 
Thus, for the models shown in Fig.~\ref{fig:bolometric_luminosity_H1000}, a duration of $\sim 10$--$30\,t_{\rm cross}$ corresponds to $\sim 10^5$--$10^6\,t_g$, and the same extent in $t/t_{\rm cross}$ corresponds to a longer physical duration for lower collision velocities.

Across most of the explored parameter space, the bolometric luminosity reaches peak values from $0.1$ up to $10\,L_{\rm Edd,BH}$. These values should be interpreted with caution, since radiative feedback is not included in the hydrodynamics evolution, and radiation forces could reduce the accretion rate and emergent luminosity in the most luminous cases. However, we note that luminosities of a few times the Eddington luminosity are not necessarily inconsistent with radiation-regulated accretion flows. The Eddington limit is obtained by balancing gravity against a spherically symmetric radiative flux, whereas the post-collision flow is highly non-spherical and radiates from extended shocked structures. The following simple estimate illustrates this point. If the shocked gas is radiation-pressure supported, then \begin{equation}
\frac{a_{\rm rad}T^4}{\rho r}\sim \frac{Gm_{\rm BH}}{r^2}.
\end{equation}
The corresponding diffusive flux over a scale $\sim r$ is
\begin{equation}
F\sim \frac{c \,a_{\rm rad}T^4}{\kappa\rho r}
\sim \frac{Gm_{\rm BH}c}{\kappa r^2},
\end{equation}
i.e. of order the local Eddington flux. The integrated luminosity is then
\begin{equation}
L\sim A_\mathrm{s} F \sim \frac{A_\mathrm{s}}{4\pi r^2}L_{\rm Edd,BH},
\end{equation}
where $A_\mathrm{s}$ is the effective area from which photons escape. Therefore, an extended non-spherical emitting region can yield an integrated luminosity above $L_{\rm Edd,BH}$ without requiring the local radiative flux to exceed the Eddington value, provided that $A_\mathrm{s} \gtrsim 4\pi r^2$. Consistently with this interpretation, two-dimensional GR radiation-hydrodynamics simulations of BHL accretion \citep{zanotti2011general} found that, although radiation feedback can reduce the accretion rate substantially, the luminosity is still of order $L/L_{\rm Edd}\sim1$--$7$. In these simulations the luminosity is produced in extended shocked structures similar to those in our scenario. In this sense, the super-Eddington luminosities in Fig.~\ref{fig:bolometric_luminosity_H1000} are not necessarily unphysical, although they should be interpreted with caution, and their precise values will require self-consistent radiation-hydrodynamics simulations.

Returning to Fig.~\ref{fig:bolometric_luminosity_H1000}, we find that the most luminous models, corresponding to relatively low velocities and low/intermediate surface densities, sustain luminosities of several $L_{\rm Edd,BH}$. In several cases, the late phase is as luminous as, or more luminous than, the early phase, despite the overall decline in density, because radiation from the inner hot region becomes progressively less trapped as the optical depth decreases. In contrast, the faintest models, mainly at $v=0.2c$, remain  $\lesssim 0.1\,L_{\rm Edd,BH}$ because the intercepted gas mass decreases steeply with velocity ($m_a \propto v^{-4}$; Sec.~\ref{sec:physical-setup}). The light curves can exhibit short-timescale fluctuations that are unlikely to correspond to observable variability and should therefore not be overinterpreted. %

The light curves show a diversity of temporal morphologies. To aid the discussion, we group the most representative temporal behaviors into four broad classes, based mostly on the relative luminosity of the early and late phases. This classification is intended only as a descriptive guide rather than a sharp taxonomy. The main purpose of the classification is to connect the light-curve morphology with the underlying physical evolution, in particular with the change in dominant emission sites. While the classification is clearer for representative examples, some may show intermediate or ambiguous morphologies. Broadly, we identify the following four types:  
\begin{itemize}
    \item \emph{Type I: Plateau.} Sustained emission with modest secular evolution. This behavior is seen when the early escaping radiation from the BHL shock is followed relatively smoothly by emission from the inner convective region, because photon trapping is not strong enough to produce an interruption between the two components.
    \item \emph{Type II: Fade out.} This behavior is favored at higher velocities, where the available mass reservoir $m_a\propto \Sigma v^{-4}$ is smaller. In these cases, the BHL-shock-powered early emission fades rapidly and the later convective component remains comparatively weak.
    \item \emph{Type III: Delayed turn-on.} The late emission becomes stronger than the early emission. This behavior is associated with low velocities and low/intermediate densities, for which the gas supply is larger than in Type I, so initially radiation escape is suppressed but the emission becomes increasingly visible as the optical depth decreases. The late rise is associated mainly with the gradual emergence of radiation from the inner convective region.
    \item \emph{Type IV: Dip-and-rebrightening.} A bright early phase followed by a pronounced minimum and a late-time rebrightening. This morphology is most prominent at high $\Sigma$, where the early escaping emission is dominated by comparatively outer, lower-$\tau$ regions of the BHL shock, while radiation from the inner convective region is initially trapped and only becomes visible after sufficient optical-depth decline.
\end{itemize}

With the caveat mentioned above in mind, two qualitative morphological trends are evident in Fig.~\ref{fig:bolometric_luminosity_H1000}. First, increasing $v$ favors quickly fading light curves: faster impacts intercept less mass ($m_a \propto v^{-4}$) and produce weaker, more rapidly fading transients. Second, increasing $\Sigma$ favors dip-and-rebrightening behavior. At high surface density, the optical depth is large and the early emission is dominated by the outer portions of the BHL shock. This outer shock fades before the optical depth has decreased sufficiently for radiation from the inner region to escape. An intermediate phase then occurs in which the outer shock has dimmed while radiation from the inner region remains trapped, producing a pronounced luminosity dip. Once the optical depth drops further, emission from the inner region becomes visible and the system rebrightens. A similar trend will also be present in the band-limited light curves and the spectral evolution discussed in Sec.~\ref{sec:specific_luminosity_lightcurves}.

\subsubsection{Total radiated energy}
In addition to luminosities, it is useful to consider the total radiated energy,
\begin{equation}
E_{\rm rad} = \int L_\mathrm{bol}(t)\,dt.
\end{equation}
In practice, this integral is evaluated over the simulation interval, $1 < t/t_{\rm cross} < 30$\footnote{The post-processing is applied only to the simulation data at $t>t_{\rm cross}$, i.e., once the black hole has collided with the disc.}. For the most optically thick models, the reported $E_{\rm rad}$ may be underestimated because the luminosity has not fully declined by the end of the simulation interval.

\begin{figure}
    \centering
    \includegraphics[width=\columnwidth]{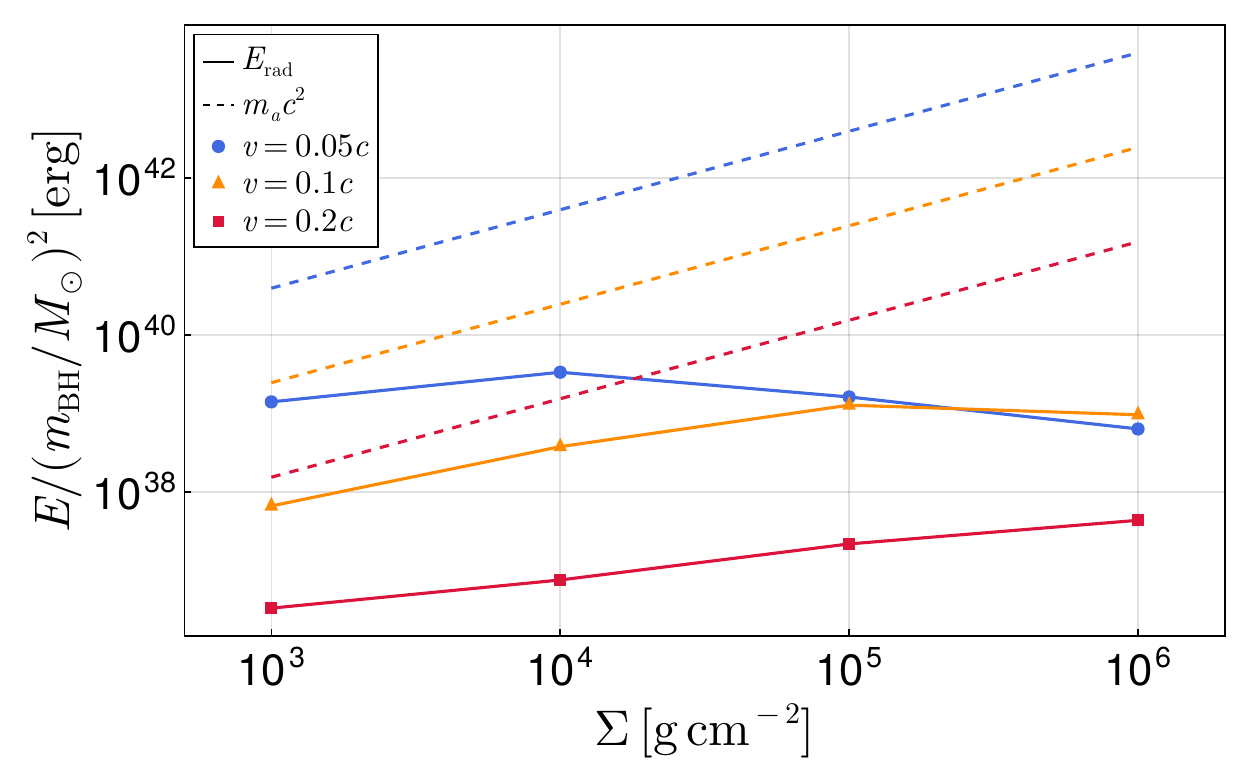}
    \vspace{-3mm}
    \caption{   
\textbf{Total radiated energy for $H=1000\,r_g$.}
Total radiated energy $E_{\rm rad}=\int L(t)\,dt$, normalized by $(m_{\rm BH}/M_\odot)^2$, shown as a function of disc surface density $\Sigma$ for three values of the collision velocity $v$. For each parameter set, the figure also shows the corresponding intercepted rest-mass energy scale $m_a c^2$ (dashed lines). 
Both quantities are normalized in the same way and are shown for direct comparison of surface density and collision velocity.
}
\label{fig:total_radiated_energy}
\end{figure}

Fig.~\ref{fig:total_radiated_energy} shows $E_{\rm rad}$, normalized by $(m_{\rm BH}/M_\odot)^2$, for various velocities and disc densities. The total radiated energy shows a strong dependence on the collision velocity $v$, decreasing by up to two orders of magnitude between $v=0.05c$ and $0.2c$ at fixed surface density. For the slowest encounters, peak values reach $E_{\rm rad}\sim\text{few}\times10^{39}(m_{\rm BH}/M_\odot)^2\,\mathrm{erg}$, while typical energies are reduced to $\sim\text{few}\times10^{38}(m_{\rm BH}/M_\odot)^2\,\mathrm{erg}$ at $v=0.1c$ and to $\sim10^{37}\text{–}10^{38}(m_{\rm BH}/M_\odot)^2\,\mathrm{erg}$ at $v=0.2c$. Again, this trend is due to the shorter interaction times and smaller interaction length scale at higher velocities. At fixed velocity, the dependence on surface density is non-monotonic. This is because increasing $\Sigma$ has two competing effects: a larger gas reservoir, but stronger photon trapping. At low and intermediate velocities, $E_{\rm rad}$ peaks at intermediate density, $\Sigma\sim10^{4}$--$10^5\,\mathrm{g\,cm^{-2}}$, while at the highest velocity the total radiated energy varies more weakly with $\Sigma$ and peaks at $\Sigma=10^6~{\rm g~cm^{-2}}$.

The hydrodynamics simulations in L25 find that the total accreted mass onto the secondary across parameter space is within $\sim 40\%$ of $m_a=\pi r_a^2\Sigma$ (see the right panel of Fig.~4 in L25). The main energy source for radiation is ultimately the binding energy of the accreted matter with respect to the secondary. This motivates defining an estimate of the radiative efficiency as
\begin{equation}
\eta_{\rm rad} := \frac{E_{\rm rad}}{m_a c^2}.
\end{equation}
As shown in Fig.~\ref{fig:total_radiated_energy}, we find $E_{\rm rad} \ll m_a c^2$ for all parameter combinations explored. The implied efficiencies are small but non-negligible, typically $\eta_{\rm rad}\sim10^{-3}\text{--}10^{-2}$, with the largest values obtained for the slowest encounters and lowest surface densities, and a significant decrease towards higher densities, due to optically thicker flows suppressing radiation escape.

\subsection{Spectral dependence}
\label{sec:results_temp}
Even though bolometric-luminosity light curves provide insight into the time-dependent energetics of the collision, they can be misleading for observational interpretation. Detectability is inherently frequency-dependent, and therefore a meaningful comparison with observations requires computing the specific luminosity, $\nu L_\nu$, in relevant electromagnetic bands. Therefore, we now turn to spectral diagnostics of the emission. In Sec.~\ref{subsec:sed}, we show time-averaged spectral energy distributions (SEDs), and in Sec.~\ref{sec:specific_luminosity_lightcurves}, we discuss the specific-luminosity light curves at various observing photon frequencies, showing the spectral evolution across parameter space.

\subsubsection{Spectral energy distributions}
\label{subsec:sed}
\begin{figure*}
    \centering
    \includegraphics[width=0.9\textwidth]{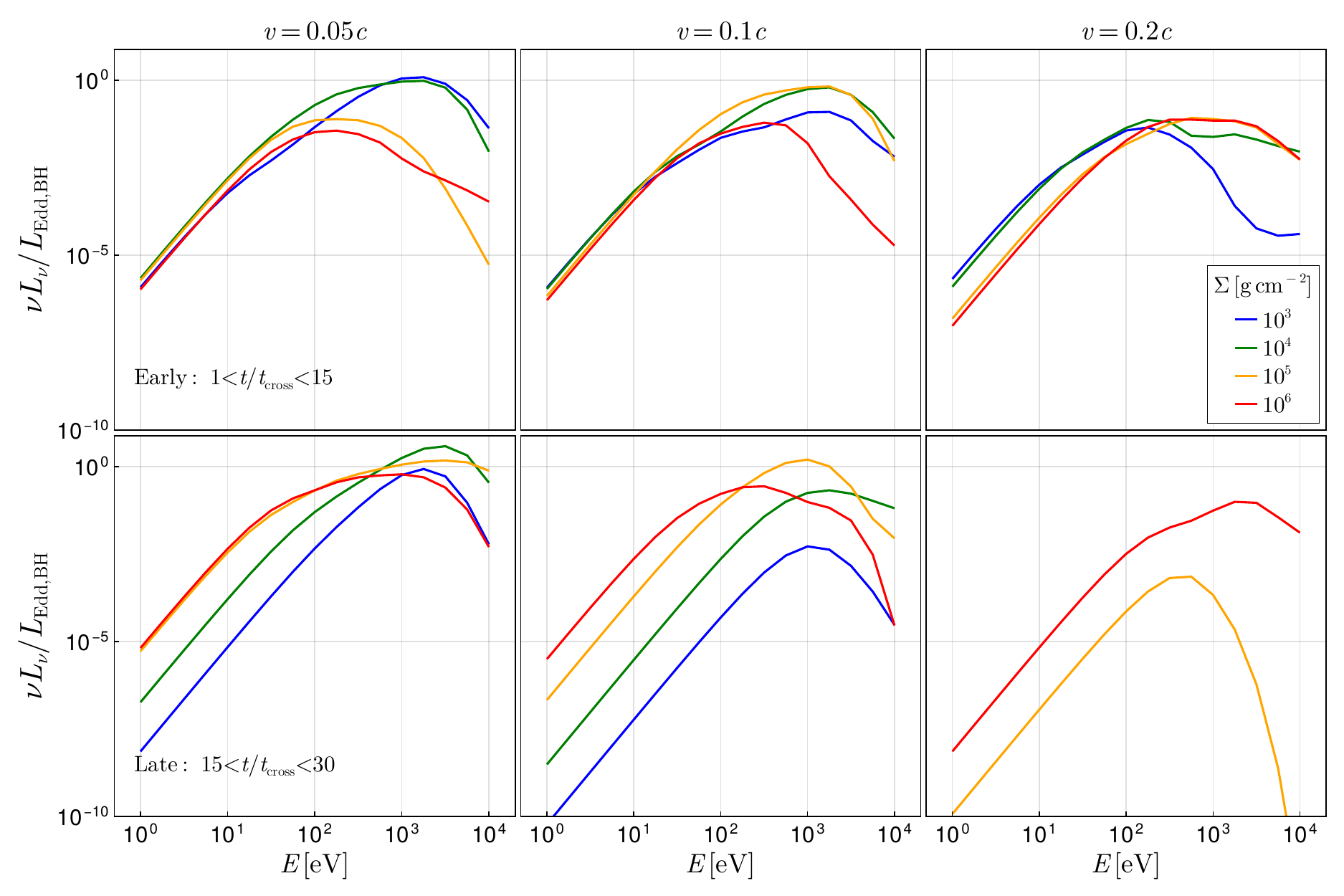}\\[0.6em]
    \caption{
    \textbf{Time-averaged SEDs for $m_{\rm BH}=10^4M_\odot$.}
    Columns correspond to black hole velocities $v=0.05c$, $0.1c$, and $0.2c$, while the top and bottom panels show the early and late phases, respectively.
    Colors denote the disc surface density $\Sigma$. The emission is generically dominated by the soft X-ray band. The curves for $\Sigma = 10^3$ and $10^4~{\rm g~cm^{-2}}$ at $v=0.2c$ in the late phase are not shown because the gas becomes very optically thin, invalidating the assumptions underlying our local temperature estimate. 
    }
    \label{fig:sed_mosaic}
\end{figure*}

\begin{figure*}
    \centering
    \includegraphics[width=\textwidth]{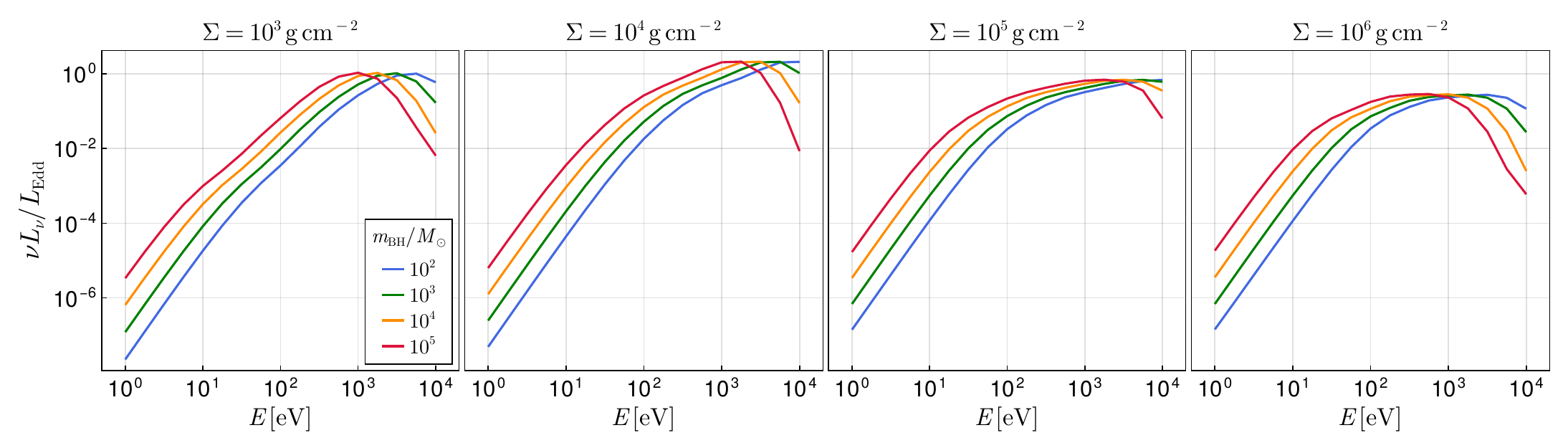}\\[0.6em]
    \caption{
    \textbf{Time-averaged SEDs for $v=0.05c$, varying black hole mass.}
    Time-averaged spectral energy distributions $\nu L_\nu/L_{\rm Edd,BH}$ shown as a function of photon energy $E$. Columns correspond to disc surface densities $\Sigma=10^3$, $10^4$, $10^5$, and $10^6~{\rm g~cm^{-2}}$. Colors denote the black hole mass $m_{\rm BH}$.
    }
    \label{fig:sed_mass_dependence}
\end{figure*}

The emission spectra produced by the collision are typically dominated by soft X-rays and exhibit dependence on collision velocity, disc surface density, and evolution phase. We first consider time-averaged SEDs, by splitting the simulation time interval in two halves: an early phase corresponding to $1 < t/t_{\rm cross} < 15$, and a late phase, $15 < t/t_{\rm cross} < 30$. Fig.~\ref{fig:sed_mosaic} shows the time-averaged SEDs, expressed as $\nu L_\nu/L_{\rm Edd,BH}$, for a representative black hole mass of $m_{\rm BH} = 10^4M_\odot$ and a range of black hole velocities and disc surface densities. The time-averaged SEDs are shown as a function of photon energy for both the early and late phases. 

The spectra typically rise from optical/UV energies to a broad peak in the soft X-ray band. The peak is typically at $E\sim1\,\mathrm{keV}$, with a high-energy tail extending toward $E\sim10\,\mathrm{keV}$. At the highest surface densities and $v \lesssim 0.1c$, however, the early-phase spectrum is softer and peaks around $E\sim100\,\mathrm{eV}$.

Some broad spectral trends are visible. For $v=0.05c$ and $0.1c$, larger $\Sigma$ suppresses the high-energy tail above the keV peak, whereas low-$\Sigma$ models retain substantial emission up to $\sim10\,\mathrm{keV}$. In several models, particularly at high $\Sigma$ and low velocity, the late-phase emission is harder than the early-phase emission. At $v=0.2c$, the early spectra do not show the same ordering with surface density, while a similar comparison in the late phase is prevented by the models becoming optically thin.

The late-time hardening and the surface-density trend at $v=0.05c$ and $0.1c$ can be understood as follows. The characteristic diffusion time scales as $t_{\rm diff} \sim \tau\,\ell_{\rm esc}/c$, and, at fixed geometry, higher $\Sigma$ implies larger optical depth and thus longer $t_{\rm diff}$. When $t_{\rm diff}$ is longer, photons remain coupled to the gas long enough to thermalize efficiently, producing a softer spectrum and suppressing the hard tail. Conversely, lower disc density, higher collision velocity (which reduces the gas reservoir and optical depth), and later times (as the flow is accreted and diluted) all reduce $t_{\rm diff}$, allowing photons to escape with less thermalization and yielding a harder spectrum within our prescription. Perhaps counterintuitively, the late-time hardening is not inconsistent with the progressive accretion of the gas, but rather caused by the reduction in $t_{\rm diff}$, since the thermalization efficiency decreases as the gas is accreted, and the overall density and optical depth decrease.

This interpretation applies while the emitting regions remain within the domain of our temperature prescription. As the flow becomes optically thin, these regions are excluded from the spectral integral, as specified in Sec.~\ref{sec:specific_luminosities}. Consequently, the tendency toward harder emission with decreasing optical depth cannot be extrapolated through this transition. This restriction is particularly relevant to the $v=0.2c$ models and helps explain why they do not follow the same spectral ordering as the slower encounters. The trends discussed here refer to the modeled optically thick component. An additional optically thin contribution, particularly at high energies, cannot be constrained by this calculation.

In Fig.~\ref{fig:sed_mass_dependence} we show time-averaged SEDs over the whole interval $1 < t/t_{\rm cross} < 30$, varying black hole mass in the IMBH regime for $v=0.05c$ and various disc densities. The SEDs soften with increasing secondary mass, with the emission generically dominated by soft X-rays within the IMBH range. The effect is the same at other collision velocities, and it can be understood in terms of the scalings described at the end of Sec.~\ref{sec:parameter_space}: while the local solution to Eq.~\eqref{eq:T_implicit} does not depend on $m_{\rm BH}$, the lower bound scales as $T_{\rm BB} \propto m_{\rm BH}^{-1/4}$, shifting part of the emission to higher photon energies for lower-mass secondaries, which explains the mild spectral trend we observe here.  

\subsubsection{Specific luminosity light curves}
\label{sec:specific_luminosity_lightcurves}
\begin{figure*}
    \centering
    \includegraphics[width=0.9\textwidth]{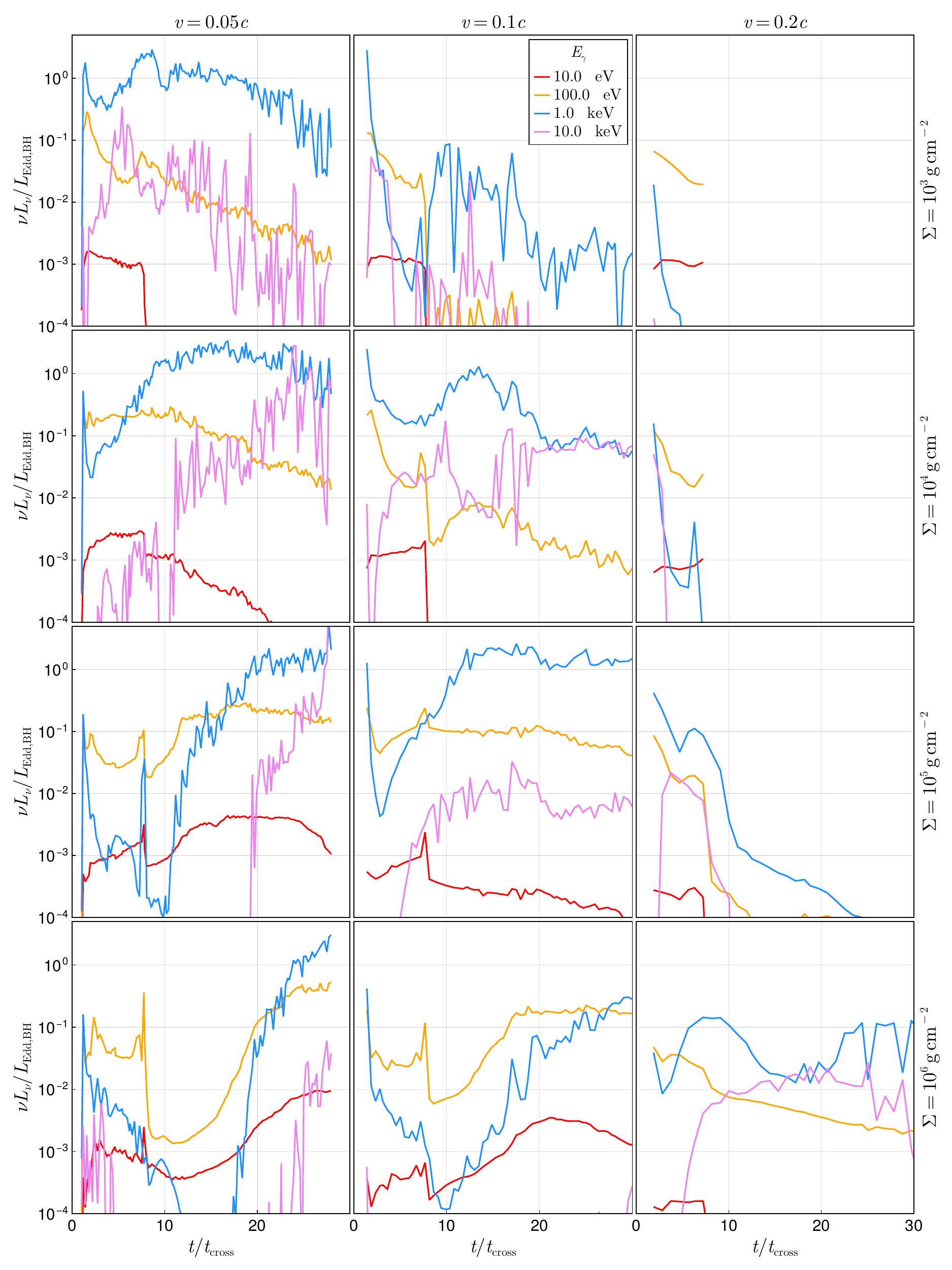}
    \caption{
    \textbf{Specific luminosity light curves for $m_{\rm BH}=10^4M_\odot$.} Specific luminosity $\nu L_\nu/L_{\rm Edd,BH}$ shown as a function of time for simulations with $H=1000\,r_g$. Columns correspond to black hole velocities $v=0.05c$, $0.1c$, and $0.2c$, while rows correspond to disc surface densities $\Sigma = 10^3$, $10^4$, $10^5$, and $10^6~{\rm g~cm^{-2}}$. Colors denote the observing energies, $E = 10$, $10^2$, $10^3$, and $10^4~{\rm eV}$. Because the horizontal axis is normalized by $t_{\rm cross}=2H/v$, the same extent in $t/t_{\rm cross}$ corresponds to a longer physical duration for lower collision velocities. Short-timescale fluctuations in the curves are unlikely to correspond to observable variability and should not be overinterpreted.}
    \label{fig:specific_lightcurve_mosaic}
\end{figure*}

Fig.~\ref{fig:specific_lightcurve_mosaic} shows the specific luminosity light curves $\nu L_\nu$ for $m_{\rm BH}=10^4\,M_\odot$ at $E=10$, $10^2$, $10^3$, and $10^4$~eV. We take the $10^4\,M_\odot$ case as representative of the IMBH range because the spectrum depends only weakly on $m_{\rm BH}$ within the range of $10^2$--$10^5M_\odot$ (see Fig.~\ref{fig:sed_mass_dependence} and the scalings discussed in Sec.~\ref{sec:parameter_space}). %

The emission is generically soft X-ray (0.1--1 keV) dominated, consistent with the time-averaged SEDs. For $\Sigma=10^3$--$10^4~\mathrm{g\,cm^{-2}}$ and $v \le 0.1c$, the $1$~keV light curve typically reaches the highest peak luminosity and the $100$~eV component is typically subdominant. The $10$~eV (optical/UV) emission is suppressed by one to two orders of magnitude relative to the soft X-ray band and is therefore unlikely to be a good observational target (see Sec.~\ref{subsec:thindisc}). The $10$~keV emission is also strongly subdominant in all models.

For higher surface densities, $\Sigma \gtrsim 10^5~\mathrm{g\,cm^{-2}}$, there is a more noticeable spectral evolution compared to the lower density models. In these cases the early emission is relatively softer, with the $100$~eV band dominating during the initial phase. At later times, however, the $1$~keV band rebrightens as the effective optical depth decreases and the inner, hotter region becomes visible. In several high-$\Sigma$ models, the peak at $1$~keV occurs significantly later than that at $100$~eV, indicating a hardening of the spectrum during the flare evolution.

\section{Discussion}
\label{sec:discussion}

In this section, we interpret our results in terms of the conditions expected in astrophysical scenarios. First, we discuss some useful generic properties of the local collision problem: the duration of the accretion-powered flare, and its possible superposition with a prompt ejecta-powered component. Second, we connect the local quantities to global disc--orbit parameters by mapping the collision velocity and disc surface density to a Keplerian orbit intersecting a standard thin accretion disc. This mapping provides a way to translate our simulation results into qualitative flare regimes in the plane of recurrence period and primary mass, and to estimate in which observing bands the flare can be visible above the global disc emission. Third, we apply these results to observed systems, discussing the extent to which black hole--disc collisions could account for the known population of QPEs, and what they predict for the case of OJ~287. We end the section with a discussion of detectability prospects.   

\subsection{Flare-duration scaling}
\label{sec:flare_duration_scaling}
A key question for connecting our light curves to observations is how the characteristic flare duration scales with the secondary mass and the collision parameters. Our results show that the accretion-powered emission typically persists for $10$--$30\,t_{\rm cross}$, i.e. $\sim 10^5$--$10^6\,t_g$, depending on the model (Sec.~\ref{subsec:bolometric_luminosities}). A simple estimate of the flare duration can be obtained by modeling the post-collision phase as the depletion of a bound reservoir of shocked gas. Let the mass of gas that becomes available to feed the post-collision accretion flow scale with the intercepted mass,
\begin{equation}
m_{\rm res}\sim f_{\rm res}\,m_a, 
\end{equation}
where $f_{\rm res}$ is the fraction of the intercepted gas that remains bound (L25 indicates that this fraction is of order unity). Let the characteristic accretion rate during the flare be a fraction $f_{\dot m}$ of the characteristic mass flux,
\begin{equation}
\dot m_{\rm acc}\sim f_{\dot m}\,\dot m_a,
\end{equation}
Then the characteristic depletion time is
\begin{equation}
t_{\rm flare}\sim \frac{m_{\rm res}}{\dot m_{\rm acc}}
\sim \frac{f_{\rm res}}{f_{\dot m}}\,\frac{m_a}{\dot m_a}
\sim \chi\,\frac{r_a}{v}
=2\chi\left(\frac{v}{c}\right)^{-3} t_g,
\label{eq:tflare_scaling}
\end{equation}
where $\chi:=f_{\rm res}/f_{\dot m}$.

We can calibrate $\chi$ to our light curves using the characteristic duration of the emission for $v=0.05c$, for which $t_{\rm flare}\sim 30\,t_{\rm cross}\sim10^6\,t_g$. This yields $\chi\sim60$. With this normalization, Eq.~(\ref{eq:tflare_scaling}) becomes
\begin{equation}
t_{\rm flare}\approx 10^6\,t_g\,v_{0.05}^{-3}\,\chi_{60},
\label{eq:tflare}
\end{equation}
where $\chi_{60}:=\chi/60$, or, in physical units,
\begin{equation}
t_{\rm flare}\approx 14~{\rm h}\,m_{\rm BH,4}v_{0.05}^{-3} \chi_{60}.
\label{eq:tflare_cgs}
\end{equation}
Thus, for IMBH secondaries the duration of the accretion-powered episode is of order hours to days, with a steep sensitivity to the collision velocity. 
The depletion timescale may not represent the true flare duration if it is shorter than the average diffusion timescale. However, we find that the depletion timescale is longer than the diffusion timescale for $\Sigma \lesssim 10^5~{\rm g~cm^{-2}}$, which is the expected range in most astrophysical scenarios. For the moment, we adopt the depletion timescale as the duration, implicitly assuming that the density is low enough. A more detailed treatment of the flare duration including higher density will be necessary in future work including radiation hydrodynamics simulations.

\subsection{Relation to ejecta-powered emission and multi-component flares}
\label{sec:multi_component_flares}
The unbound ejecta produced during the collision may power electromagnetic emission as it expands and cools \citep[][L25]{lehto1996oj, ivanov1998hydrodynamics, franchini2023quasi, linial2023emri+}. Therefore, black hole--disc collisions can give rise to structured, multi-peaked flares produced by the superposition of the long-lived accretion-powered component modeled above and the prompt radiation from the unbound ejecta. We do not compute the ejecta-powered light curve directly in this work, because doing so would require following the unbound material beyond the local simulation domain. Instead, we rely on simple analytical estimates as described below. 

Based on Eq.~(29) of L25 for the peak timescale of ejecta emission, and adopting the estimate from Sec.~\ref{sec:flare_duration_scaling} for the accretion-powered flare duration, the ejecta peak time is typically earlier than the accretion-powered flare duration by a factor of
\begin{equation}
\frac{t_{\rm peak, ej}}{t_{\rm flare}} \sim 0.06\,\Sigma_4^{1/2}v_{0.05}^{1/2}\left(\frac{v_{\rm ej}}{v}\right)^{-1/2}\left(\frac{m_{\rm ej}}{0.1 m_a}\right)^{1/2}\chi_{60}^{-1},
\end{equation}
where $m_{\rm ej}$ and $v_{\rm ej}$ are the characteristic ejecta mass and velocity, respectively, scaled to the typical values measured in the simulations. The peak luminosity of the ejecta is estimated in L25 as $L_{\rm peak, ej} \sim L_{\rm Edd,BH}$, essentially independent of parameters. Therefore, the observed flare may consist of a brief precursor from the ejecta followed by a broader accretion-powered component. In some regimes, the accretion-powered emission alone already produces two maxima, such that the addition of an ejecta precursor may yield a three-component structure.

\begin{figure*}
\centering
\includegraphics[width=\textwidth]{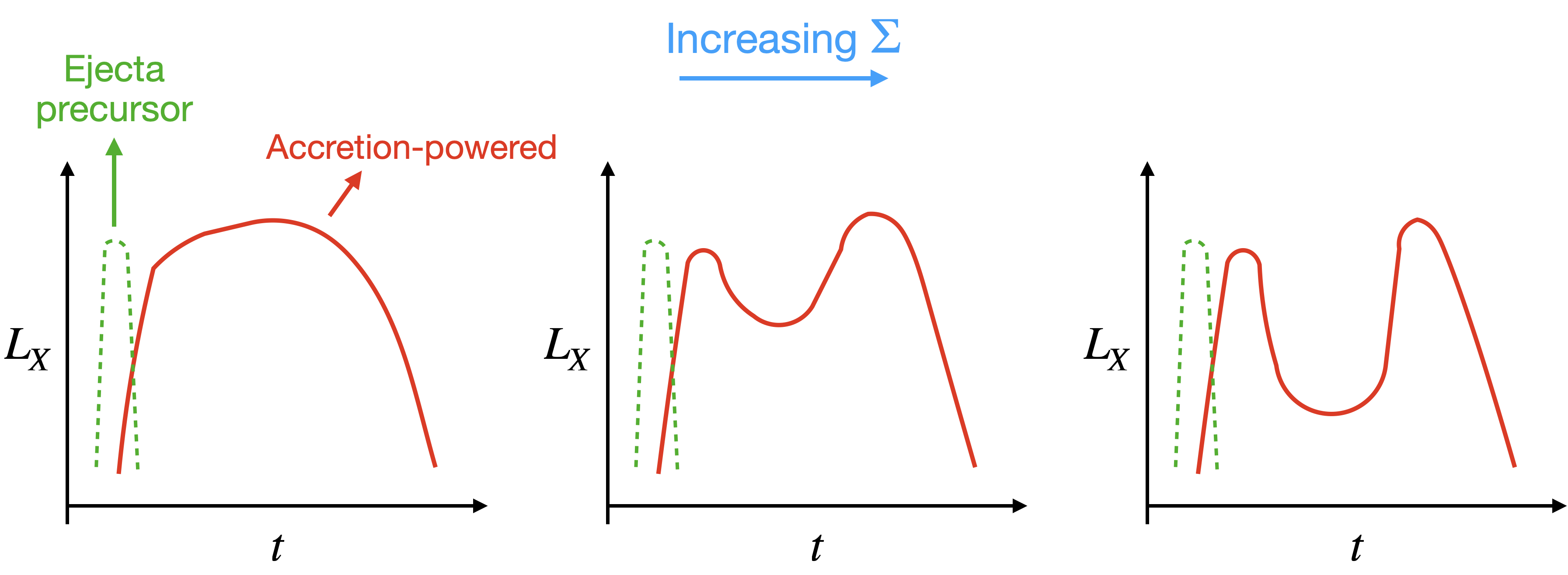}
\vspace{-3mm}
\caption{
Schematic illustration of structured flare morphologies from black hole--disc collisions at relatively low velocities, $v \lesssim 0.05c$. 
The dashed green curve represents a prompt ejecta-powered precursor, while the solid red curves illustrate the accretion-powered emission. The curves are illustrative and not derived from a quantitative light-curve model.
}
\label{fig:multi_component_flare}
\end{figure*}

Fig.~\ref{fig:multi_component_flare} provides a schematic representation of this phenomenology at relatively low encounter velocities ($v \lesssim 0.05c$), where the accretion-powered emission is strongest, and increasing surface density enhances dip-and-rebrightening morphology. The intermediate-time minimum deepens and the late-time maximum becomes more pronounced as $\Sigma$ increases, consistent with stronger photon trapping and delayed emergence of radiation from the inner convective region. An interesting implication of this morphological trend is that if the disc surface density evolves significantly over a timescale in which the disc-impactor system remains active, the individual flare morphology may evolve accordingly. In Sec.~\ref{sec:qpes}, we speculate on this possibility for QPE sources in the aftermath of TDEs.

\subsection{Mapping the predicted flares to astrophysical systems}
\label{subsec:thindisc}
To connect the parameter space explored in this work to astrophysical systems, we relate the secondary black hole velocity $v$ to the Keplerian orbital velocity and the disc surface density $\Sigma$ to the properties of a standard thin accretion disc \citep{shakura1973black} around the primary. This mapping allows us to relate the flare duration to the recurrence period of the emission and to compare the flare luminosity to that of the underlying global disc. 
\subsubsection{Orbital parameters and flare duration}
We define the QPE recurrence period as $P_{\rm QPE}:=P_{\rm orb}/2$, where $P_{\rm orb}$ is the orbital period, assuming a circular orbit. This corresponds to the interval between successive disc crossings, and assumes that each crossing produces a visible flare. For a circular Keplerian orbit around a primary of mass $M_c$, the orbital radius can be expressed as
\begin{equation}
R = \left(G M_c P_{\rm QPE}^2/\pi^2 \right)^{1/3} \approx 314\,R_g\, M_{\mathrm{c,6}}^{-2/3} P_{\mathrm{QPE,d}}^{2/3},
\end{equation}
where $R_g:=GM_c/c^2$ is the gravitational radius of the primary, $M_{\mathrm{c,6}}:=M_c/10^6\,M_\odot$, and $P_{\mathrm{QPE,d}}:=P_{\mathrm{QPE}}/1\,\mathrm{d}$. The corresponding orbital velocity of the secondary is
\begin{equation}
v_{\rm K} = \left(GM_c/R\right)^{1/2} \approx 0.056c\,M_{\mathrm{c,6}}^{1/3}P_{\mathrm{QPE,d}}^{-1/3}.
\label{eq:velocity}
\end{equation}
Combining Eq.~\eqref{eq:velocity} with the relation $t_{\rm flare}\propto m_{\rm BH} v^{-3}$ from Eq.~\eqref{eq:tflare_cgs}, we obtain 
\begin{equation}
\frac{t_{\rm flare}}{P_{\rm QPE}} \approx 0.4\, q_{-2} \,\chi_{60}.
\label{eq:tdur_PQPE}
\end{equation}
where $q:= m_{\rm BH}/M_c$ is the mass ratio, and $q_{-2}:=q/10^{-2}$.  Here we have assumed that the relative velocity is $v = v_{\rm K}$. For an inclined orbit relative to a Keplerian disc, the relative velocity can differ from $v_{\rm K}$ by a geometric factor. This factor is generically of order unity, except for nearly coplanar, prograde encounters, where it can become very small. Our model, based on simulations of axisymmetric, orthogonal collisions, remains appropriate for generic disc-crossing orbits, such as those of EMRIs originating from stellar systems with roughly isotropic angular momentum distribution. However, it cannot capture the nearly coplanar regime where the collision properties depend sensitively on the inclination angle. In this low-relative velocity regime, the accretion radius is expected to increase, and thus the flares may be more energetic than in our setup \citep{franchini2023quasi, liu2026quasi}.
\subsubsection{Thin-disc surface density scalings}
The disc surface density at the crossing radius depends on the local pressure regime of the disc. In the outer regions of a Shakura--Sunyaev disc\footnote{We only consider the regimes where the opacity is dominated by electron scattering, neglecting the outermost region where free--free opacity dominates.} the pressure is dominated by gas, yielding \citep{frank2002accretion}
\begin{equation}
\Sigma_{\rm (gas)} \approx 1.2\times10^5
\,M_{\mathrm{c,6}}^{3/5}
P_{\mathrm{QPE,d}}^{-2/5}
\dot m_{-1}^{3/5}
\alpha_{-1}^{-4/5}
\eta_{-1}^{-3/5}
\,\mathrm{g\,cm^{-2}},
\label{eq:sigma_gas}
\end{equation}
where $\dot M_{\rm Edd,c}:= L_{\rm Edd,c}/(\eta c^2)$, $L_{\rm Edd,c}$ is the Eddington luminosity of the primary, $\eta$ is the radiative efficiency of the disc, $\dot m:=\dot M_c/\dot M_{\rm Edd,c}$, $\dot m_{-1}:=\dot m/0.1$, $\alpha_{-1}:=\alpha/0.1$, and $\eta_{-1}:=\eta/0.1$. In the innermost, radiation-pressure dominated region the surface density instead scales as
\begin{equation}
\Sigma_{\rm (rad)} \approx 1.1\times10^5
\,M_{\mathrm{c,6}}^{-1}
P_{\mathrm{QPE,d}}
\dot m_{-1}^{-1}
\alpha_{-1}^{-1}
\eta_{-1}
\,\mathrm{g\,cm^{-2}}.
\label{eq:sigma_rad}
\end{equation}
\subsubsection{Band-dependent visibility against the global disc emission}
To assess in which bands the flare may be visible above the emission of the primary disc, we compare the simulated specific luminosities with a simple estimate of the thin-disc emission. Writing the global disc luminosity as
\begin{equation}
L_{\rm disc}\approx \dot m\,L_{\rm Edd,c}
\approx 10\,\dot m_{-1}\,q_{-2}^{-1}\,L_{\rm Edd,BH},
\label{eq:Ldisc}
\end{equation}
where $L_{\rm Edd,c}$ is the Eddington luminosity of the primary, and the characteristic inner-disc temperature as
\begin{equation}
k_{\rm B} T_{\rm disc}\sim k_{\rm B} \left(\frac{3GM_c\dot M_c}{8\pi\sigma R_{\rm ISCO}^3}\right)^{1/4} \approx 30\text{--}60\,{\rm eV}\,\dot m_{-1}^{1/4}M_{c,6}^{-1/4},
\end{equation}
we can obtain order-of-magnitude estimates for the disc specific luminosity at different photon energies. Near the thermal peak of the thin disc ($10$--$100$~eV), we can write
\begin{equation}
\frac{\nu L_{\nu, \rm flare}}{\nu L_{\nu, \rm disc}}
\approx \frac{\nu L_{\nu, \rm flare}}{L_{\rm disc}}
\approx 0.1\left(\frac{\nu L_{\nu, \rm flare}}{L_{\rm Edd,BH}}\right) \dot{m}_{-1}^{-1}  q_{-2}\,,
\label{eq:visibility}
\end{equation}
where we approximated $\nu L_{\nu, \rm disc} \approx L_{\rm disc}$ and inserted Eq.~\eqref{eq:Ldisc}. The ratio $\nu L_{\nu, \rm flare} / L_{\rm Edd,BH}$, which ultimately determines the visibility, depends on frequency, and its typical values can be read from the specific-luminosity light curves in Fig.~\ref{fig:specific_lightcurve_mosaic}.

At $E\sim10$ eV, which lies near the thermal peak of the global-disc emission, our models typically give $\nu L_{\nu,{\rm flare}}(10\,{\rm eV})\lesssim10^{-3}\,L_{\rm Edd,BH}$. The flare is therefore expected to be strongly outshone in the optical/UV band for all astrophysically relevant parameters (see Eq.~\eqref{eq:visibility}). Note that this does not exclude the possibility of a delayed UV counterpart, as recently discovered \citep{guo2026evidence} in the QPE source ``Ansky'' \citep{sanchez2024sdss1335+, hernandez2025discovery, chakraborty2025rapidly, zhu2025ultraviolet, hernandez2025nicer, chakraborty2026positive, chakraborty2026relativistic}, if the direct X-ray flare is reprocessed elsewhere, for example in the outer regions of a warped disc.

At $E\sim100$~eV, also near the characteristic disc temperature, the simulated flares reach $\nu L_{\nu}(100\,{\rm eV})\sim 0.1 L_{\rm Edd,BH}$, implying that direct visibility in this band may be possible only for $\dot{m}_{-1} \lesssim 0.01 q_{-2}$, i.e., for example, for $q\approx10^{-2}$ and $\dot{m} \lesssim 10^{-3}$. 

At $E\sim1$ keV the global-disc emission is deep in its Wien tail ($E/kT_{\rm disc}\sim20$--$30$) and is therefore exponentially suppressed. On the other hand, the flare typically reaches $\nu L_{\nu, \rm flare}(1~{\rm keV})\sim 0.1$--$10\,L_{\rm Edd,BH}$. Consequently, the accretion-powered flare should be most readily detectable in the soft X-ray band even when its bolometric luminosity is smaller than that of the global disc.
\subsubsection{Qualitative flare regimes in the \texorpdfstring{$(P_{\rm QPE},M_c)$}{(PQPE,Mc)} plane}
\begin{figure*}
    \centering
    \includegraphics[width=0.9\textwidth]{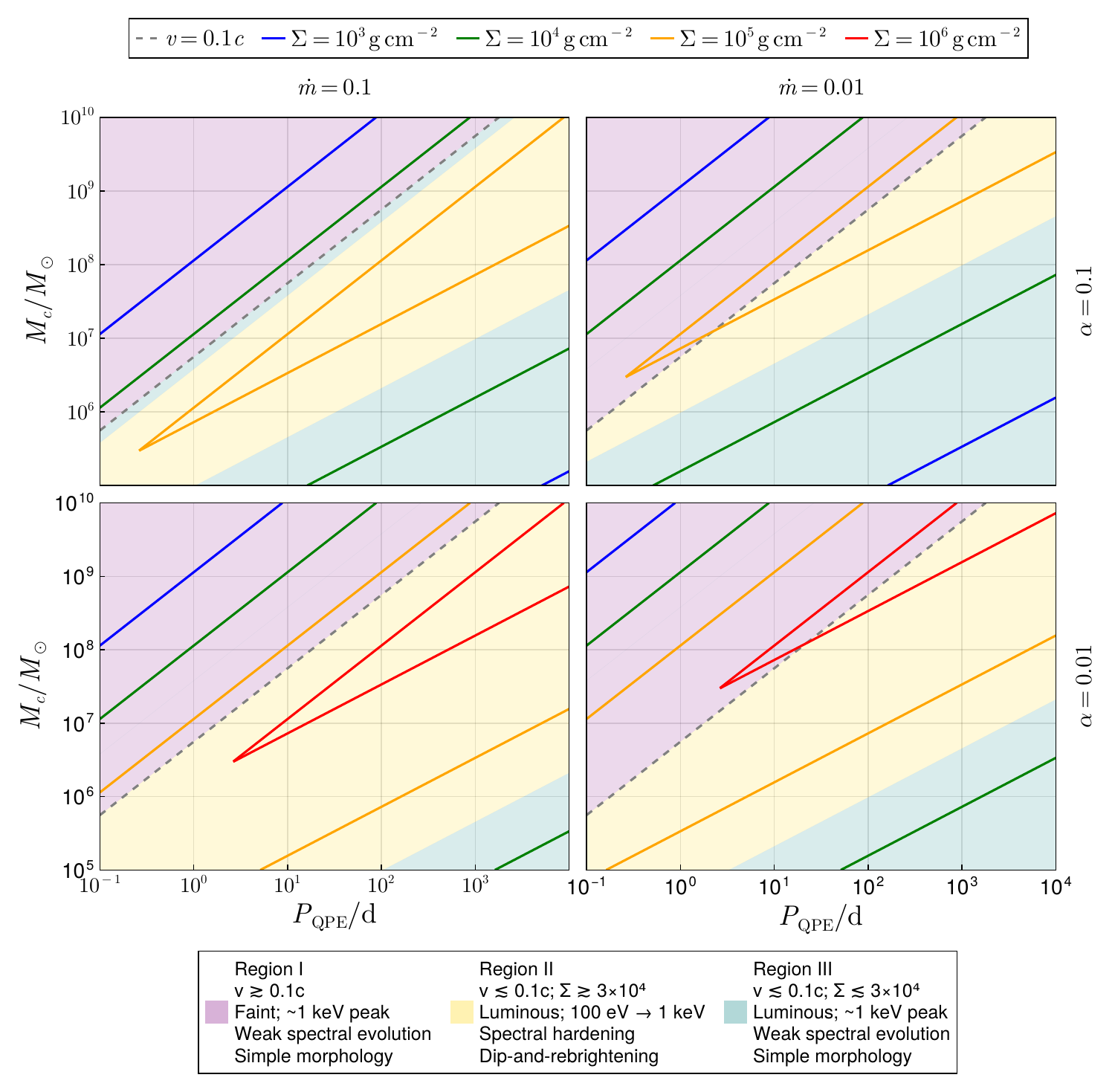}
\caption{
\textbf{Regions of parameter space with qualitatively different flare properties.}
The panels show the plane of QPE recurrence period $P_{\rm QPE}$ and primary mass $M_c$ for representative values of the accretion rate $\dot{m}=0.1$ and $0.01$ and viscosity parameter $\alpha=0.1$ and $0.01$, assuming a radiative efficiency $\eta=0.1$. The dashed curve corresponds to a Keplerian velocity $v=0.1c$ for a circular orbit of period $P_{\rm orb}=2P_{\rm QPE}$. Colored curves show contours of constant disc surface density $\Sigma$ derived from the Shakura--Sunyaev thin-disc scalings in the gas- and radiation-pressure dominated regimes. The change of slope of the contours reflects the transition between these regimes. 
Shaded regions indicate the approximate parameter ranges where the simulations predict different qualitative emission behaviors (see Sec.~\ref{subsec:thindisc}).
}
    \label{fig:thin-discs}
\end{figure*}
Fig.~\ref{fig:thin-discs} shows the parameter space explored in this work in the plane of $P_{\rm QPE}$ and $M_c$, for representative values $\dot m=0.1$ and $0.01$, $\alpha=0.1$ and $0.01$, and assuming $\eta=0.1$. Based on our results, we identify three broad physical regimes in parameter space, defined by the two quantities that most strongly control the flare properties: the collision velocity $v$, which sets the intercepted gas mass and therefore the available energy reservoir, and the disc surface density $\Sigma$, which primarily regulates photon trapping and spectral evolution. The qualitative labels assigned in Fig.~\ref{fig:thin-discs} are based on the light curves shown in Fig.~\ref{fig:bolometric_luminosity_H1000} (for the overall energetics), and Fig.~\ref{fig:specific_lightcurve_mosaic} (for the rough spectral evolution). The chosen boundaries are not intended as sharp physical transitions, but rather as approximate guides to the qualitative behaviors inferred from our results. This description refers only to the accretion-powered component modeled in this work. The schematic ejecta-powered precursor discussed in Sec.~\ref{sec:multi_component_flares} is not considered in this classification. Concretely, we identify the following qualitative regions:

\begin{itemize}

\item \textbf{Region I:} ($v \gtrsim 0.1c$). At high collision velocities the accretion radius is small and the intercepted gas mass is strongly reduced (because $m_a\propto v^{-4}$). As a result, the flares are generally faint and short-lived. Their light curves have typically simple morphologies\footnote{Except at the very highest densities $\Sigma \approx 10^6~{\rm g~cm^{-2}}$ where dip-and-rebrightening behavior starts to be found. For simplicity, we do not consider an additional class for this case.}, with spectra peaking around $1$~keV and showing little spectral evolution. See, for example, the model with $(v, \Sigma) = (0.2c, 10^5~\rm{g~cm^{-2}})$.

\item \textbf{Region II}: ($v \lesssim 0.1c$ and $\Sigma \gtrsim 3\times10^4~\mathrm{g\,cm^{-2}}$). Here the gas reservoir remains large, but the optical depth is sufficiently high that photon trapping strongly affects the observed emission. The flares can still be luminous, but their early emission is softer, often peaking near $\sim 100$~eV, and the spectrum hardens with time as radiation from the inner hotter region becomes visible. The light curves show dip-and-rebrightening behavior. See, for example, the model with $(v, \Sigma) = (0.05c, 10^5~\rm{g~cm^{-2}})$.

\item \textbf{Region III}: ($v \lesssim 0.1c$ and $\Sigma \lesssim 3\times10^4~\mathrm{g\,cm^{-2}}$). In this regime the secondary intercepts a large gas reservoir, while the optical depth is moderate enough that radiation can escape relatively promptly. The resulting flares are typically luminous, again with simple morphology, peaking around $\sim 1$~keV and with little spectral evolution. See, for example, the model with $(v, \Sigma) = (0.05c, 10^3~\rm{g~cm^{-2}})$.

\end{itemize}

In Fig.~\ref{fig:thin-discs}, the dashed curve corresponds to $v=0.1c$ obtained from Eq.~(\ref{eq:velocity}) assuming a circular Keplerian orbit. It approximately separates a high-velocity regime, where flares are faint, from a slower-interaction regime, where flares are luminous. Colored curves show contours of constant surface density $\Sigma$ derived from the thin-disc scalings of Eqs.~\eqref{eq:sigma_gas} and \eqref{eq:sigma_rad}. Because $\Sigma$ scales on $(M_c,P_{\rm QPE})$ with opposite signs of their exponents in the gas- and radiation-pressure regimes (compare Eqs.~\eqref{eq:sigma_gas} and \eqref{eq:sigma_rad}), the constant-$\Sigma$ contours change slope across the transition between these branches, producing the wedged shapes visible in Fig.~\ref{fig:thin-discs}. The shaded regions illustrate Regions~I, II and III in pink, yellow, and cyan, respectively.

The conditions defining each regime can be recast in terms of $P_{\rm QPE}$ and $M_c$. For example, the defining condition of Region~I ($v\gtrsim0.1c$) corresponds to
\begin{equation}
\text{(Region I)}\quad
P_{\mathrm{QPE,d}} \lesssim 0.17\,M_{\mathrm{c,6}} .
\end{equation}
Since the surface-density contours change slope across the transition between the dominant source of pressure, the $\Sigma$ conditions cannot in general be written as a simple scaling of $P_{\rm QPE}$ and $M_c$. However, for typical parameters $\dot m\sim0.01$--$0.1$ and $\alpha\sim0.01$--$0.1$, and within the mass and period range shown in Fig.~\ref{fig:thin-discs}, the II/III transition lies entirely within the gas-pressure dominated regime. In that case, we can use Eq.~(\ref{eq:sigma_gas}) to give the conditions
\begin{equation}
\begin{aligned}
\text{(Region II)}\quad
&P_{\mathrm{QPE,d}} \gtrsim 0.17 M_{\mathrm{c,6}}, \\
&P_{\mathrm{QPE,d}} \lesssim32\,M_{\mathrm{c,6}}^{3/2}\dot m_{-1}^{3/2}\eta_{-1}^{-3/2}\alpha_{-1}^{-2}.
\end{aligned}
\end{equation}
\begin{equation}
\begin{aligned}
\text{(Region III)}\quad
&P_{\mathrm{QPE,d}} \gtrsim 0.17 M_{\mathrm{c,6}}, \\
&P_{\mathrm{QPE,d}} \gtrsim 32\,M_{\mathrm{c,6}}^{3/2}\dot m_{-1}^{3/2}\eta_{-1}^{-3/2}\alpha_{-1}^{-2}.
\end{aligned}
\end{equation}

\subsection{Implications for known QPE sources}
\label{sec:qpes}
An interesting feature of the depletion-time estimate is that, once the collision velocity is related to the orbital period through Eq.~\eqref{eq:tdur_PQPE}, it yields $t_{\rm flare}\propto P_{\rm QPE}$ assuming that $v < 0.1c$ and $\Sigma \lesssim 10^4\,\mathrm{g\,cm^{-2}}$. Therefore, for a class of black hole--disc collisions within that range of velocities and densities, the model gives a scaling with recurrence period consistent with the observed trend in known QPEs. For primaries of order $10^7\,M_\odot$ and secondaries in the range $m_{\rm BH}\sim$ few $\times10^4\,M_\odot$, the estimate gives $t_{\rm flare}/P_{\rm QPE}\sim0.05$--$0.2$, comparable to the typical QPE duty cycle. A caveat to this argument is that the scaling derived in Sec.~\ref{sec:flare_duration_scaling} assumes that the flare duration is set by the depletion time of the bound gas reservoir. Additionally, the parameter $\chi$ may not be constant, in which case the scaling $t_{\rm flare} \propto P_{\rm QPE}$ would be modified.

Energetic and population arguments nevertheless indicate that black hole--disc collisions may be unlikely to account for the majority of QPEs. For stellar-mass black holes, these encounters may be energetically insufficient \citep{linial2023emri+}, except possibly for nearly-coplanar configurations where the relative velocity is significantly lower and therefore the interaction radius is larger \citep{franchini2023quasi, liu2026quasi}. On the other hand, if the bulk of X-ray QPEs was powered by IMBH secondaries, the inferred event rate \citep{arcodia2024cosmic, chakraborty2025discovery} would imply the existence of a large population of massive secondaries in galactic nuclei, leading to such rapid SMBH growth by mergers with IMBHs that the sole observation of $\sim10^6\,M_\odot$ SMBHs would be rendered implausible.  In our work, we find total radiated energies $E_{\rm rad}\sim10^{37}\text{--}10^{39}\,(m_{\rm BH}/M_\odot)^2\,\mathrm{erg}$ across our parameter space, comparable to the estimates in \citet{linial2023emri+} on which these arguments are based. Thus, our results are broadly consistent with them. 

However, these arguments do not rule out the possibility that a subset of QPEs, or other repeating transients yet to be identified, could be powered by such collisions. Moreover, as mentioned above, nearly coplanar, stellar-mass-black-hole collisions may circumvent these difficulties. Unfortunately, we cannot draw any conclusions on this particular case because our framework, based on simulations of orthogonal collisions, although suitable for generic inclinations, is unreliable in the nearly coplanar regime, where the flare properties depend very sensitively on the inclination angle.

In light of the discussion in Sec.~\ref{sec:multi_component_flares}, particularly interesting tests may come from sources that show evidence for multiple flare components. In this respect, eRO-QPE1 \citep{arcodia2021x} appears especially interesting for the black hole--disc collision scenario. This source shows evidence for complex eruptions that can be interpreted as partially superposed flares \citep{arcodia2022complex}, and recent long-term monitoring suggests that EMRI--disc collision models are broadly consistent with its observed timing properties \citep{chakraborty2024testing}. For its recurrence time of $\sim 18.5$~h and a central mass in the $\sim10^5$--$10^6\,M_\odot$ range, Eq.~\eqref{eq:velocity} gives a Keplerian velocity $v\sim0.04$--$0.06c$, well within the slow-collision regime where our models predict the most luminous and longest-lived accretion-powered flares. Moreover, the observed duty cycle, $t_{\rm flare}/P_{\rm QPE}\approx 7.6/18.5\approx0.4$, is broadly consistent with Eq.~\eqref{eq:tdur_PQPE} for a mass ratio $q\sim10^{-2}$, corresponding to an IMBH secondary. In this sense, the characteristic timescale and luminosity scale of eRO-QPE1 are consistent with the black hole--disc collision picture presented here. However, the long-term variations and irregular overlap of eruptions reported in this source \citep{chakraborty2024testing} suggest that additional ingredients beyond the local approximation of our framework may be required, including orbital precession and the global structure of the disc.

More extreme structured QPEs such as the recently discovered in J2344 \citep{baldini2026discovery}, five years after a TDE-like flare \citep{homan2023discovery, goodwin2024radio}, are harder to accommodate in the black hole--disc collision picture. J2344 exhibits broad flares accompanied by a crest of narrower, hotter flares, an unprecedented phenomenology within the current QPE sample. For a recurrence time of $\sim 12$~h and a central mass $M_c\sim 10^7\,M_\odot$, Eq.~\eqref{eq:velocity} gives $v\approx 0.15c$, toward the high-velocity end of the parameter space explored here, where the intercepted gas mass is strongly reduced and the flares tend to be fainter and shorter-lived. Matching the observed hour-scale durations and luminosity scale would require a secondary of at least a few $\times 10^4\,M_\odot$ (unless, perhaps, the collisions are nearly coplanar). The corresponding gravitational-wave inspiral time (Eq.~\eqref{eq:tgw}) would be only of order $\sim 10$--$90$~yr. Observing a system in this short-lived configuration may therefore be unlikely. However, there is evidence that J2344 may be a recently faded AGN, which could enhance both TDE and EMRI rates \citep{jiang2025embers}. In such environments, the probability of detecting a system with this short lifetime may be significantly increased.

Recent spectroscopy of Ansky has provided constraints on the mass expelled during its eruptions \citep{chakraborty2026relativistic}. The comparison with the mass required to power the radiation through accretion depends on the assumed radiative efficiency. The efficiencies obtained here, \(\eta_{\rm rad}\sim10^{-3}\)–\(10^{-2}\), imply an accreted mass approximately \(10\)–\(100\) times larger than the efficiency of \(0.1\) used in \citet{chakraborty2026relativistic}, for the same bolometric radiated energy. Such low efficiencies alleviate the tension raised there for accretion-powered eruption models, which was based on an excess of expelled over accreted mass. A comparison with the observed outflow velocities and spectral evolution would require further modelling.

Finally, we comment on an interesting implication of the flare-morphology dependence with disc density, mentioned at the end of Sec.~\ref{sec:multi_component_flares} and schematically illustrated in Fig.~\ref{fig:multi_component_flare}. As discussed there, an evolving disc surface density could produce a corresponding evolution of the morphology of individual collision flares. A natural setting in which this could occur is in a post-TDE disc. A number of QPEs has been detected following a previous TDE, or another episode of nuclear activation, in the same nucleus \citep{chakraborty2021possible, quintin2023tormund,nicholl2024quasi,hernandez2025discovery,chakraborty2025discovery,bykov2025further}, suggesting that $\sim 10 \%$ of TDEs may be followed by QPEs. This connection can be explained in the so-called ``EMRI+TDE'' picture \citep{linial2023emri+}. In this picture, some time after the TDE, a compact disc is formed which gradually spreads outwards \citep[see e.g.][]{mummery2020spectral,mummery2024fundamental}. If an EMRI companion is already present, QPE-like activity may begin once the expanding disc reaches the orbital radius of the secondary. The surface density encountered by the secondary would then rise rapidly and subsequently decline as the finite mass reservoir is accreted and spread over larger radii on the viscous timescale, typically of order years.

In such a scenario, one may speculate that the flare morphology evolves together with the local surface density sampled by the secondary. At early times, when the density at the crossing radius is higher, the soft X-ray flares may be more structured, with multiple temporal peaks. As the density decreases, the flares may transition toward smoother, more monotonic profiles with weaker spectral evolution. This speculation is only qualitative, since making quantitative predictions and comparing with known QPE sources would require coupling the local collision problem studied here to the global viscous evolution of a realistic post-TDE disc. Moreover, for high mass ratios to the primary, $q\gtrsim 0.1$, the secondary may deplete the post-TDE disc too rapidly by repeated impacts \citep{mummery2025collisions}, therefore challenging this picture altogether for very high mass ratios.

\subsection{Implications for OJ~287}
\label{sec:oj287}
OJ~287 is one of the best-known candidates for an SMBH binary in which a secondary may periodically impact the accretion disc of a much more massive primary. In the standard impact model \citep[e.g.][]{lehto1996oj, valtonen2008massive}, the primary mass is $M_c\sim10^{10}M_\odot$ and the secondary mass is $\sim10^{8}M_\odot$, with an orbital period of $\sim9$ yr (rest frame) and high eccentricity ($e\sim0.6$--$0.7$). The circular Keplerian estimate of Eq.~\eqref{eq:velocity} gives $v \approx 0.07c$. 

The observed optical outbursts of OJ~287 reach luminosities of order $L\sim10^{46}$--$10^{47}\,\mathrm{erg\,s^{-1}}$. In our models the luminosity scale typically reaches $\sim1$--$10\,L_{\rm Edd,BH}$, implying a secondary mass $m_{\rm BH} \sim 10^{8}\,M_\odot$, consistent with the masses invoked in binary-SMBH interpretations of OJ~287. However, combining this mass estimate with the flare-duration scaling derived in Sec.~\ref{sec:flare_duration_scaling}, $t_{\rm flare}\approx 14~{\rm h}\, m_{\rm BH,4} v_{0.05}^{-3}$, gives a characteristic duration $t_{\rm flare}\sim$ few years for $m_{\rm BH}\sim10^{8}M_\odot$ and $v\sim0.07c$. This is substantially longer than the observed weeks-to-months duration of the major optical outbursts in OJ~287. This tension is in line with previous arguments in \citet{linial2023emri+}. In the extreme case of nearly coplanar collisions, the required secondary mass could be lower, potentially predicting flare durations closer to the observed values \citep{franchini2023quasi, liu2026quasi}.

\subsection{Detectability and observational prospects}
\begin{figure*}
    \centering
    \includegraphics[width=\textwidth]{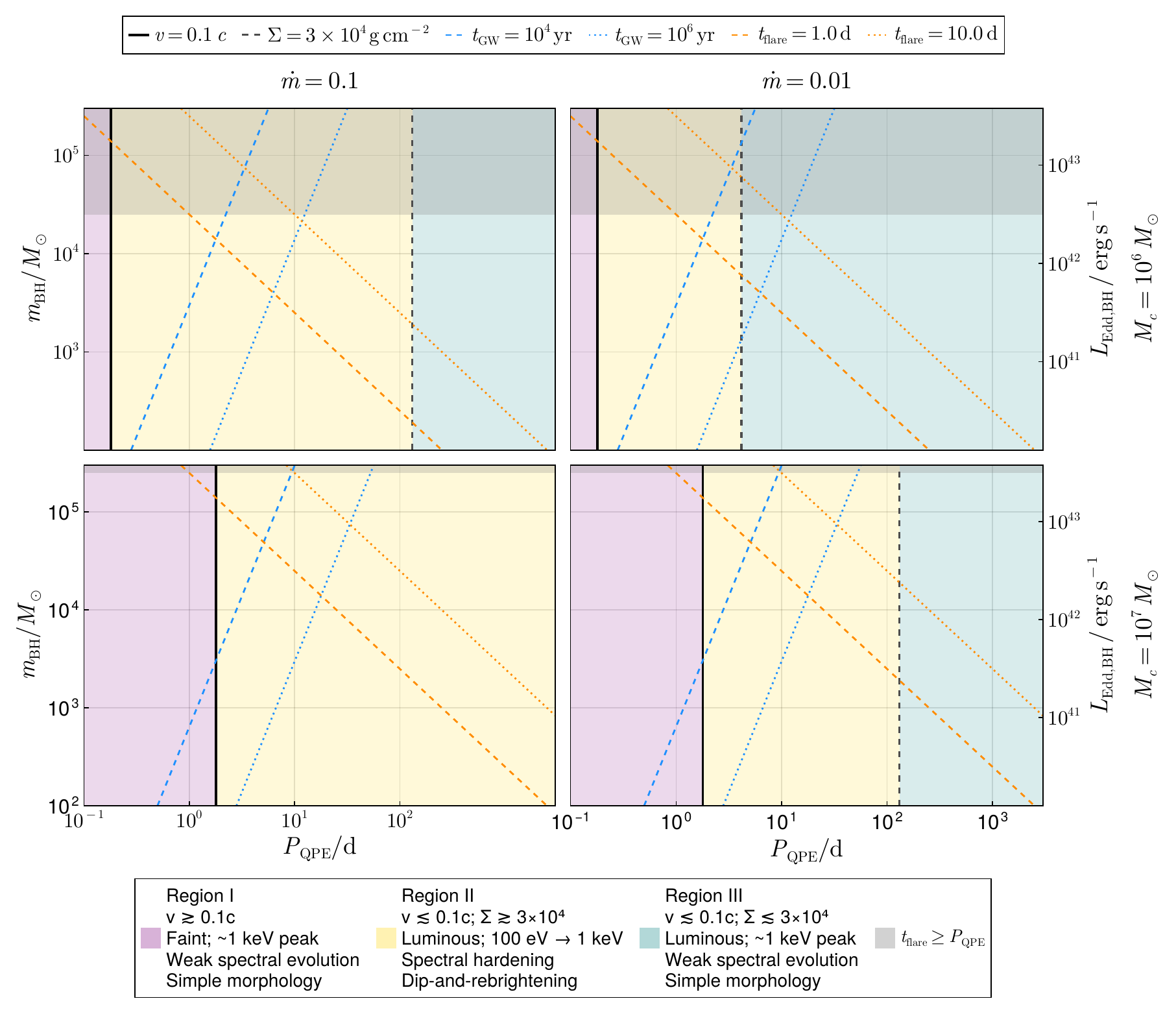}
    \vspace{-3mm}
    \caption{ \textbf{Parameter space in the $(P_{\rm QPE},\,m_{\rm BH})$ plane.} The panels show the recurrence period $P_{\rm QPE}$ and secondary black-hole mass $m_{\rm BH}$ for representative values of the primary mass $M_c$ and accretion rate $\dot m$, assuming the Shakura--Sunyaev disc scalings of Sec.~\ref{subsec:thindisc}, with $\alpha = 0.05$ and $\eta = 0.1$.  Colored regions indicate the three qualitative flare regimes defined from Fig.~\ref{fig:thin-discs}. Colored curves show contours of flare duration $t_{\rm flare}$ from the analytic estimate of Eq.~\eqref{eq:tflare_cgs}, while blue curves show contours of gravitational-wave inspiral time $t_{\rm GW}$. The gray shaded regions mark where $t_{\rm flare} \geq P_{\rm QPE}$, implying the emission is not QPE-like and our framework does not apply.}
    \label{fig:mbh-mosaic}
\end{figure*}

Combining the thin-disc mapping of Sec.~\ref{subsec:thindisc} with the flare-duration estimate of Sec.~\ref{sec:flare_duration_scaling} allows us to identify where QPE-like black hole--disc collision flares are most likely to be observable in parameter space. In Fig.~\ref{fig:mbh-mosaic}, we show a diagram in the $(P_{\rm QPE},m_{\rm BH})$ plane, complementary to Fig.~\ref{fig:thin-discs}. The figure shows contours of $t_{\rm flare}$ from Eq.~\eqref{eq:tflare}, as well as contours of the gravitational-wave inspiral timescale of the binary system \citep[e.g.,][]{shapiro2024black},
\begin{equation}
t_{\rm GW}
\approx
\frac{5}{256}\,\frac{c^5 R^4}{G^{3} M_c^2 m_{\rm BH} }
\approx 3\times10^3\,{\rm yr}\,m_{{\rm BH},4}^{-1}M_{{\rm c},6}^{-2/3}P_{{\rm QPE},d}^{8/3},
\label{eq:tgw}
\end{equation}
Systems in which the disk is a post-TDE disk and the inspiral timescale of the EMRI is shorter than $\sim 10^4~{\rm yr}$ are unlikely to be found, since the intrinsic rate of TDEs is estimated as $\sim 10^{-4}$--$10^{-5}\,{\rm yr^{-1} galaxy^{-1}}$. At fixed $M_c$, increasing $P_{\rm QPE}$ lowers the orbital velocity and therefore increases the intercepted gas mass, favoring more luminous flares. Increasing $m_{\rm BH}$ raises the luminosity scale but also lengthens the flare duration and shortens the gravitational-wave inspiral time. The most favorable systems are therefore expected in an intermediate region of parameter space: outside the high-velocity regime $v\gtrsim0.1c$, and not so far toward large $m_{\rm BH}$ that successive flares overlap or the inspiral time becomes very short.

A useful guide is provided by Eq.~\eqref{eq:tdur_PQPE}, $t_{\rm flare}/P_{\rm QPE}\approx 0.4\,q_{-2}\chi_{60}$. When this ratio is small, individual impacts can produce well-separated QPE-like flares. As the ratio approaches unity, consecutive flares may overlap, leading instead to lower-contrast quasi-periodic or sustained variability. Our results do not allow us to describe that regime reliably, because they are based on a local approximation and do not include the global structure of the disc or the orbit of the secondary. In this sense, the most favorable QPE-like candidates are expected in Regions~II--III with moderate mass ratios, producing luminous, temporally distinct flares, while Region~I sources should generally be fainter and therefore harder to detect unless they are unusually nearby. Most of Region~I corresponds to inspiral timescales shorter than $10^4~\rm{yr}$, and therefore is unlikely to be realized at least in post-TDE systems. 

Our results also indicate that wide-field soft-X-ray monitoring is the natural discovery channel for these events. In the IMBH regime, the flares are expected to last hours to days and to peak mainly in the $\sim0.1$--$1$~keV band. Einstein Probe \citep{yuan2022einstein} is particularly useful for discovering such events, whereas XMM--Newton  \citep{jansen2001xmm} is well suited to pointed follow-up. Among future prospects, {\it NewAthena} \citep{barcons2017athena} should further improve the follow-up of these systems, while the wide-field mission concept THESEUS \citep{amati2021theseus} may also become relevant for discovery if realized.

\section{Conclusions}
\label{sec:conclusions}
In this work we studied the electromagnetic signatures of black hole--disc collisions by applying a post-processing framework of radiative processes to relativistic hydrodynamics simulations. Our main conclusions are as follows:
\begin{itemize}
\item The observable emission is dominated by the long-lived, post-collision accretion flow onto the secondary black hole, rather than by cooling of the unbound ejecta. The luminosity of this accretion-powered emission is broadly around $L_{\rm Edd,BH}$, and can reach several times this value in the most luminous cases.
\item The emission is generically dominated by soft X-rays, typically peaking near $E\sim1\,{\rm keV}$.
\item The overall energetics depends strongly on the collision velocity because the intercepted gas mass scales as $m_a\propto v^{-4}$. Relatively lower-velocity encounters therefore produce substantially more luminous and more energetic flares, with $E_{\rm rad}\sim10^{37}\text{--}10^{39}(m_{\rm BH}/M_\odot)^2\,{\rm erg}$, over the explored parameter space.
\item The disc surface density mainly regulates photon trapping and thus the temporal and spectral evolution of the flare. High-$\Sigma$ models show softer early emission and more pronounced non-monotonic light curves, including dip-and-rebrightening behavior.
\item A depletion-time estimate calibrated to our results suggests characteristic durations of hours to days for secondaries of mass $\sim 10^4M_\odot$, with $t_{\rm flare}\propto P_{\rm QPE}$ (Eqs.~\eqref{eq:tflare_cgs} and \eqref{eq:tdur_PQPE}).  This connection is suggestive for QPE-like transients, although the scaling may not hold in the most optically thick cases ($\Sigma \gtrsim 10^5~{\rm g~cm^{-2}}$).
\end{itemize}
Our results also show some conditions under which black hole--disc collisions may produce observable transients in galactic nuclei. Because the intercepted gas mass decreases steeply with increasing collision velocity, luminous, long-duration flares are favored in relatively slow encounters, while high-velocity collisions are expected to produce weaker flares. In the IMBH regime explored here, the most promising discovery channel is wide-field soft-X-ray monitoring.

Several limitations should be kept in mind. The underlying hydrodynamics simulations neglect radiation forces and cooling, the radiative post-processing relies on simplified prescriptions for photon escape, and the calculations are local rather than global (in the sense that we do not take into account the global structure of the disc). Future simulations including radiation coupling, global disc structure, and orbital evolution will be necessary to refine these predictions. More complete radiation-hydrodynamics computations and future observations of repeating nuclear transients will be essential for testing the relevance of this scenario to QPE-like sources and related systems. Despite its limitations, the present work shows that black hole--disc collisions can naturally produce long-lived, soft-X-ray accretion-powered flares. 

\section*{Acknowledgements}
The authors thank Pietro Baldini, Kenta Hotokezaka, Christopher Irwin, Itai Linial, Tatsuya Matsumoto, and Andrew Mummery for useful discussions. This work benefited from discussions during the Yukawa Institute for Theoretical Physics (YITP) workshop YITP-T-25-02, ``Multi-Messenger Astrophysics in the Dynamic Universe''. The authors thank the YITP at Kyoto University for hospitality during the workshop. Numerical computations were performed on the Sakura cluster at the Max Planck Computing and Data Facility (MPCDF). This work was in part supported by Grant-in-Aid for Scientific Research (grant No.~23H04900) of Japanese MEXT/JSPS.
A.T.L.L. acknowledges support by NASA under award No. 80NSSC25K7213.

\section*{Software}
This work made use of the Julia programming language \citep{bezanson2017julia} and the following packages: \texttt{Makie.jl} \citep{DanischKrumbiegel2021} for visualization, \texttt{DataFrames.jl} \citep{JSSv107i04} for data handling, and \texttt{Eikonal.jl} for solving the eikonal equation for the optical depth.

\section*{Data Availability}
The data underlying this article will be shared on reasonable request to the corresponding author.

\bibliographystyle{mnras}
\bibliography{reference, scholar, software}

\appendix
\section{Shock identification}
\label{app:shock_id}
We identify shocks at cell interfaces by comparing primitive variables on the left (L) and right (R) sides of each interface. Let $v_n$ denote the velocity component normal to the interface, $\rho$ the mass density, $p$ the gas pressure, and $s = p/\rho^\Gamma$ an entropy measure, with $\Gamma$ the adiabatic index. We classify the upstream state by requiring that the normal velocity decreases across the interface from left to right\footnote{This condition holds regardless of the orientation of the shock: if the right state is upstream, $v_{n,R} < v_{n,L} < 0$; if the left state is upstream, $0 < v_{n,R} < v_{n,L}$.} and that density, pressure, and entropy increase in the downstream direction. Specifically, the left state is identified as upstream if
\begin{equation} 
\begin{aligned} 
\rho_L &< \rho_R ,\quad v_{n,L} > v_{n,R},\\ p_L &< p_R,\quad s_L < s_R, 
\end{aligned} 
\label{eq:shock_upstream_left} 
\end{equation} 
while the right state is identified as upstream if 
\begin{equation} 
\begin{aligned}
\rho_L &> \rho_R,\qquad v_{n,L} > v_{n,R},\\ p_L &> p_R,\qquad s_L > s_R. 
\end{aligned} 
\label{eq:shock_upstream_right} 
\end{equation}
Interfaces that satisfy neither set of inequalities are not classified as shocks. At cell interfaces where shocks are detected, the shock speed is
estimated using the Rankine--Hugoniot relations. Namely,
\begin{equation}
v_{\mathrm{sh}} = v_u \pm c_{s,u}\,
\sqrt{1+\frac{\Gamma+1}{2\Gamma}\left(\frac{p_d}{p_u}-1\right)},
\label{eq:vsh}
\end{equation}
with $c_{s} = \sqrt{\Gamma p/\rho}$ the sound speed. The sign is chosen so that the upstream velocity in the shock frame, $v_u-v_{\mathrm{sh}}$, points towards the interface (i.e. the plus sign when the upstream is the left state, and the minus sign when the upstream is the right state). 

\section{Photon trapping factor}
\label{app:advection_trapping}
Our advection-trapping factor is motivated by a steady toy model constructed on top of a given hydrodynamic profile, $(\rho_0,\vec{v}_0, e_0)$, where $e_0 = \rho_0\varepsilon_0$ is the internal energy density. Assuming a quasi-steady state and radiation-dominated internal energy ($\Gamma = 4/3$), the underlying hydrodynamic profile satisfies  
\begin{equation}
\nabla \cdot \left[ \left(\frac{1}{2} \rho_0 v_0^2 + \frac{4}{3}e_0 \right) \vec{v}_0  \right] = 0\,.
\label{eq:hydro_profile_1D}
\end{equation}
If we considered the effect of photon diffusion, the result would be a different profile $(\rho, \vec{v}, e)$ satisfying
\begin{equation}
\nabla \cdot \left[ \left( \frac{1}{2} \rho v^2 + \frac{4}{3} e  \right) \vec{v} - \frac{c}{3\kappa \rho} \nabla e \right] = 0\,,
\label{eq:rad_energy}
\end{equation}
where the additional term inside the divergence is the diffusive flux.
Approximating $(\rho, \vec{v}) \approx (\rho_0, \vec{v}_0)$ and using Eq.~\eqref{eq:hydro_profile_1D}, then
\begin{equation}
\nabla \cdot \left[ \left( \frac{4}{3}(e-e_0)  \right) \vec{v} - \frac{c}{3\kappa \rho} \nabla e \right] = 0\,,
\label{eq:rad_energy_2}
\end{equation}
In a one-dimensional planar problem, the equation becomes
\begin{equation}
\frac{de}{d\tau} + \frac{4v}{c}\left(e-e_0\right) = \mathrm{const.}\,,
\label{eq:rad_energy_1D}
\end{equation}
where $d\tau = -\kappa \rho dx$. In the presence of inward flow ($v<0$), the outgoing flux is exponentially attenuated by the optical depth. Therefore, we define the photon trapping factor along an escape path $\gamma_{\mathrm{esc}}$ as:
\begin{equation}
f_{\mathrm{adv}} :=
\exp\!\left[\frac{4}{c}\int_{\gamma_{\mathrm{esc}}} v^-\,\dd\tau\right],
\qquad
v^- := \min(v_r,0),
\label{eq:fadv}
\end{equation}
where $v_r$ is the radial velocity component and $\dd\tau=\kappa\rho\,\dd l$ is the differential optical depth along
$\gamma_{\mathrm{esc}}$.

\bsp	%
\label{lastpage}
\end{document}